\newif\ifTwoColumn
\newif\ifTechReport

\TechReporttrue    
									 
\ifTwoColumn
	\documentclass[journal]{IEEEtran}
	\else
		\ifTechReport
		\documentclass[journal,12pt,onecolumn,draftcls]{IEEEtran}
	\else
		\documentclass[journal,11pt,onecolumn,draftclsnofoot]{IEEEtran}
	\fi
\fi

\usepackage[utf8]{inputenc} 
\usepackage[T1]{fontenc}    
\usepackage{url}            
\usepackage{booktabs}       
\usepackage{amsfonts,amsmath,amsthm,amssymb,mathrsfs}       
\usepackage{bbm}
\usepackage{cite}
\usepackage{mathtools,xparse}
\usepackage[makeroom]{cancel}
\usepackage[acronym]{glossaries}
\usepackage{glossaries-prefix}
\usepackage{color}
\usepackage{balance}
\usepackage{enumerate}
\usepackage{algpseudocode}
\usepackage{subcaption}
\usepackage{nicefrac}       
\usepackage{microtype}      
\usepackage{xcolor}         
\usepackage{graphicx}
\usepackage[prependcaption,colorinlistoftodos]{todonotes}
\usepackage{multirow}
\usepackage{hyperref}
\usepackage{booktabs,tabularx}
\usepackage{cases}
\usepackage[ruled,vlined]{algorithm2e}

\graphicspath{{./figures/}}

\renewcommand{\qedsymbol}{$\blacksquare$}

\DeclareMathSizes{10}{10}{6}{4}

\newcommand{\continuanceref}{}

\newcommand{\R}{\mathbb{R}}

\newcommand{\N}{\mathbb{N}}

\newcommand{\mc}[1]{\mathcal{#1}}

\newcommand{\eqdef}{\coloneqq}
\newcommand{\reqdef}{\eqqcolon}

\newcommand{\col}{\mathrm{col}}

\newcommand{\set}[2]{\left\{ #1\ \left| \ #2 \right. \right\}}

\newcommand{\bs}[1]{\boldsymbol{#1}}
\newcommand{\bsone}{\boldsymbol{1}}

\newtheorem{theorem}{Theorem}[section]
\newtheorem{definition}[theorem]{Definition}
\newtheorem{proposition}[theorem]{Proposition}
\newtheorem{lemma}[theorem]{Lemma}
\newtheorem{corollary}[theorem]{Corollary}

\newtheorem{standing}[theorem]{Standing Assumption}
\newtheorem{assumption}[theorem]{Assumption}

\newcommand{\normaltext}[1]{\textnormal{#1}}

\usepackage[colorinlistoftodos]{todonotes}

\usepackage{xcolor,calc}

\newcommand{\circo}{~\raisebox{1pt}{\tikz \draw[line width=0.5pt] circle(1.3pt);}~}

\makeglossaries

\newacronym{GNEP}{GNEP}{generalized Nash equilibrium problem}
\newacronym{NEP}{NEP}{Nash equilibrium problem}
\newacronym{NE}{NE}{Nash equilibrium}
\newacronym{LQR}{LQR}{linear quadratic regulator}
\newacronym{ARE}{ARE}{algebraic Riccati equation}
\newacronym{BR}{BR}{best-response}
\newacronym{LS}{LS}{least-squares}
\newacronym{PWA}{PWA}{piecewise-affine}
\newacronym{iid}{i.i.d.\@}{independent and identically distributed}
\newacronym{wrt}{w.r.t.\@}{with respect to}
\newacronym{LTI}{LTI}{linear time-invariant}
\newacronym{mp-QP}{mp-QP}{multi-parametric quadratic program}
\newacronym{MSE}{MSE}{mean squared error}
\newacronym{EV}{EV}{electric vehicle}
\newacronym{SoC}{SoC}{state of charge}
\newacronym{DSO}{DSO}{distribution system operator}

\newglossaryentry{LP}
{
	name={LP},
	description={linear programming},
	first={\glsentrydesc{LP} (\glsentrytext{LP})},
	plural={LPs},
	descriptionplural={linear programs},
	firstplural={\glsentrydescplural{LP} (LPs)}
}

\newglossaryentry{v-GNE}
{
	name={v-GNE},
	description={variational generalized Nash equilibrium},
	first={\glsentrydesc{v-GNE} (\glsentrytext{v-GNE})},
	plural={v-GNE},
	descriptionplural={variational generalized Nash equilibria},
	firstplural={\glsentrydescplural{v-GNE} (\glsentryplural{v-GNE})}
}

\newglossaryentry{GNE}
{
	name={GNE},
	description={generalized Nash equilibrium},
	first={\glsentrydesc{GNE} (\glsentrytext{GNE})},
	plural={GNE},
	descriptionplural={generalized Nash equilibria},
	firstplural={\glsentrydescplural{GNE} (\glsentryplural{GNE})}
}

\begin{document}
	
\title{An active learning method for solving competitive multi-agent decision-making and control problems}
\author{Filippo Fabiani and Alberto Bemporad, \IEEEmembership{Fellow, IEEE}
	\thanks{This work was partially supported by the European Research Council (ERC), Advanced Research Grant
		COMPACT (Grant Agreement No. 101141351).
		The authors are with the IMT School for Advanced Studies Lucca, Piazza San Francesco 19, 55100, Lucca, Italy ({\tt \{filippo.fabiani, alberto.bemporad\}@imtlucca.it}).}}


\maketitle

\begin{abstract}
To identify a stationary action profile for a population of competitive agents, each executing private strategies, we introduce a novel active-learning scheme where a centralized external observer (or entity) can probe the agents' reactions and recursively update simple local parametric estimates of the action-reaction mappings. Under very general working assumptions (not even assuming that a stationary profile exists), sufficient conditions are established to assess the asymptotic properties of the proposed active learning methodology so that, if the parameters characterizing the action-reaction mappings converge, a stationary action profile is achieved. Such conditions hence act also as certificates for the existence of such a profile. Extensive numerical simulations involving typical competitive multi-agent control and decision-making problems illustrate the practical effectiveness of the proposed learning-based approach.
\end{abstract}


\IEEEpeerreviewmaketitle

\glsresetall

\section{Introduction}\label{sec:intro}
\IEEEPARstart{M}{\lowercase{ulti}}-agent system applications are rapidly growing 
in complexity due to the widespread use of embedded sensors and distributed control systems~\cite{astrom2011control}. 
The deployed agents can make autonomous decisions to maximize some private performance index while, at the same time, interacting with their peers. The competitive or cooperative nature of such interactions yield emerging behaviours, which in addition to the possibly large-scale structure and potential privacy issues, considerably complicate the analysis of the multi-agent system at hand. Besides, those systems made of interacting decision-makers typically feature desirable collective outcomes meant as solutions of the interaction process, which can be gathered under some (loose) stationarity notion, for example, a Nash equilibrium in noncooperative game theory \cite{facchinei2010generalized}.

The considerations above hence create a fertile ground for the problem investigated in this paper. Consider, for instance, a population of competitive agents that interact with each other through private \emph{action-reaction mappings}, which may (or may not) lead to a particular collective outcome, here identified as a \emph{stationary action profile}. By leveraging tools from the machine learning and system identification literature 
\cite{Set12,Bem23,ljung2010perspectives},
our work aims at answering the following question: given an agnostic scenario in which an external observer (or entity) is only allowed to query the action-reaction mappings, is it possible \emph{to learn} a stationary action profile of the underlying multi-agent interaction process?

\subsection{Motivating example: forecasting price-responses to control the aggregated electricity consumption in smart grids}\label{subsec:motivating_example}

Among the \emph{indirect control} strategies enabling energy flexibility offered by end-users \cite{d2022exploiting}, energy retailers or \glspl{DSO} design price signals to which price-sensitive users adjust their consumption/storage profile to meet their private needs \cite{schweppe1980homeostatic,corradi2012controlling,vespermann2017offering}.  Leveraging insights into user price-responsiveness, the price-generation process thus serves as a mechanism for managing electricity consumption. Such an indirect control method, however, rely on accurate predictions of the users' consumption responses to broadcasted price signals for activating flexibility. It is then crucial to develop tools to accomplish such a prediction task under limited knowledge about the decision process driving users' price-responsiveness. 

As an example, consider a set of $N$ \glspl{EV} populating a distribution grid \cite{ma2011decentralized,grammatico2017dynamic,cenedese2019charging}, where
every user wants to determine an optimal \gls{EV} charging schedule over a certain discrete time interval $\{1,\ldots,T\}$ by controlling the energy injection $x_i \in \R_{\ge 0}^T$ of its own \gls{EV} charger. Specifically, each user $i \in \{1,\ldots,N\}$ aims at minimizing a \emph{private} cost in the form $J_i(x_i,\sigma(\bs x)) = \|x_i\|^2_{Q_i}+c_i^\top x_i + (a (\sigma(\bs x) + d) + b \bsone_T)^\top x_i$, where $\bs x \in \R_{\ge 0}^{NT}$ represents the vector stacking all the users' decision variables, $\|x_i\|^2_{Q_i}+c_i^\top x_i$ models the battery degradation cost, and $(a (\sigma(\bs x) + d) + b \bsone_T)^\top x_i$ the electricity pricing. In particular, $\sigma(\bs x)$ denotes the \emph{aggregate demand} of the overall population of \glspl{EV}, 
usually defined as
$\sigma(\bs x) = \tfrac{1}{N} \sum_{i=1}^{N} x_i \in \R^{T}_{\ge0}$, $a>0$ represents the inverse of the price elasticity of demand, $b >0$ the baseline price, and $d\in\R^T_{\ge 0}$ the normalized average inflexible demand. In addition, each user has to satisfy both local and shared constraints due for instance to a minimum charging amount over the interval, $\bsone_T^\top x_i \ge \gamma_i \geq0$, a cap on the power injection $x_i \in [0, \bar x_i]^T$, or accounting for intrinsic grid limitations, e.g., $\sigma(\bs x)+d \in [0, \bar c]^T$. 

This formulation thus originates a so-called
generalized Nash game in aggregative form, whose solution coincides to a
Nash equilibrium \cite{facchinei2010generalized}, i.e.,
a collective charging profile, say $\bs x^\star$, in which none of the user can gain by unilaterally deviating from its current strategy. On the other hand, such an equilibrium $\bs x^\star$, which yields the aggregate consumption $\tfrac{1}{N} \sum_{i=1}^{N} x^\star_i$, heavily depends on the price signals $a$ and $b$ affecting the \emph{private} cost function of each agent, $J_i$, i.e., $\bs x^\star=\bs x^\star(a,b)$. It is then clear how a suitable design of $a$ and $b$, based on an accurate prediction of the resulting $\bs x^\star(a,b)$, allows for an efficient (for \glspl{DSO}) and profitable (for retailers) usage of the distribution grid. In \S \ref{sec:applications}, we will derive an active learning-based approach enabling an exact prediction of the outcome of the multi-agent interaction process on which little knowledge is available.

\subsection{Related work}\label{subsec:literature}
Our results can be positioned within those branches of literature referring to the learning of equilibria in an empirical game-theoretic or black-box setting.  However, our framework can encompass a broader class of multi-agent decision-making and control problems than is usually considered.

The pioneering work \cite{fearnley2013learning} focused on congestion games, showing how one can learn the agents' cost functions while querying only a portion of decision spaces. This work then originated a series of contributions investigating query and/or communication complexity of algorithms for specific classes of games \cite{babichenko2016query,goldberg2016bounds,babichenko2017communication,hart2018query}. Along the same line, in \cite{goldberg2021learning}, several schemes were devised with provable bounds on the best-response query complexity for computing approximate equilibria of two-player games with a finite number of strategies. Few other works belonging to the simulation-based game literature, instead, took a probably approximately correct learning perspective to reconstruct an analytical representation of normal form games (i.e., matrix games) for which a black-box simulator provides noisy samples of agents' utilities, devising algorithms that uniformly approximate the original games and associated equilibria with finite-sample guarantees \cite{viqueira2019learning,areyan2020improved,marchesi2020learning}.
Stochastic \cite{vorobeychik2008stochastic} or sample-average approximation \cite{vorobeychik2010probabilistic} of simulated games, and Bayesian optimization-based methods \cite{vorobeychik2008stochastic,al2018approximating,picheny2019bayesian} have also been adopted. In the former cases, the authors investigated matrix games with finite decision sets, providing an asymptotic analysis of Nash equilibria obtained from simulation-based models, along with probabilistic bounds on their approximation quality. Bayesian approaches are, instead, empirical and typically rely on Gaussian processes used as emulators of the black-box cost functions. Posterior distributions provided by the Gaussian process are successively adopted to design acquisition functions tailored to solve Nash games, which are based on the probability of achieving an equilibrium. More recently, \cite{clarke2023learning} proposed an active-set-based first-order algorithm to learn the rationality parameters of the agents taking part in a potential game from historical observations of Nash equilibria.

Suitable examples of works tackling standard problems in game theory through machine learning techniques can be found in \cite{vorobeychik2007learning,huang2022distributed,fabiani2022learning,fabiani2022personalized}. In particular, \cite{vorobeychik2007learning} addressed payoff-function learning as a standard regression problem. The methodology presented there, however, focused on games in normal form and came with no theoretical guarantees. Combining proximal-point iterations and ordinary \gls{LS} estimators, \cite{huang2022distributed} designed a distributed algorithm with probabilistic convergence guarantees to an equilibrium in stochastic games where the agents learn their own cost functions. In \cite{fabiani2022learning,fabiani2022personalized}, instead, a coordinator aims at reconstructing private information held by the agents to enable the computation of an equilibrium by designing personalized incentives affecting the cost functions.

\subsection{Summary of contributions and paper organization}
\begin{figure}[t!]
	\centering
	\ifTwoColumn
\else
	\includegraphics[width=.6\textwidth]{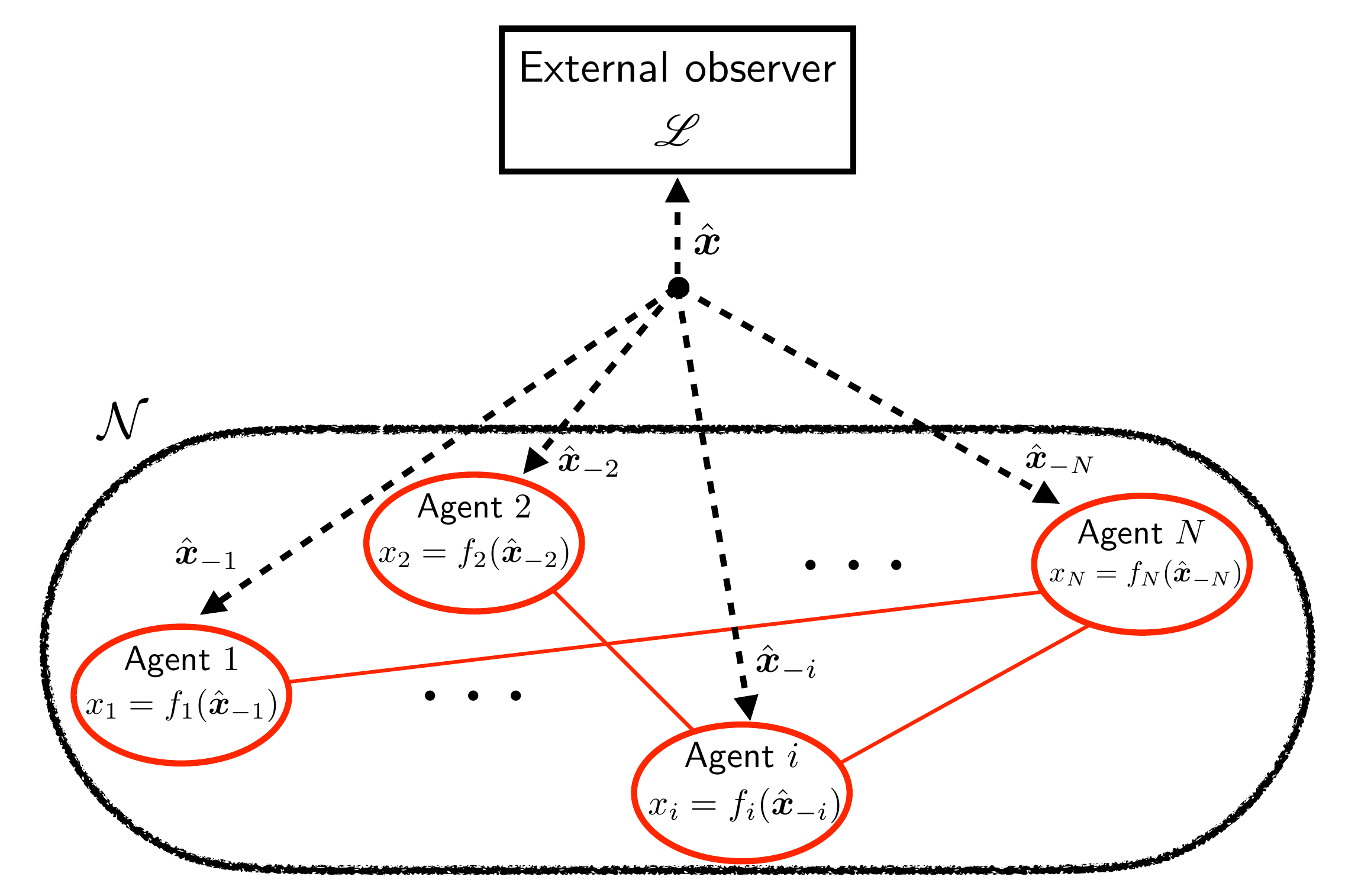}
	\fi
	\caption{An external observer with learning procedure $\mathscr{L}$ makes queries $\hat{\bs x}$ and observes each reaction to $\hat{\bs x}_{-i}$ (dashed black lines) taken by agent $i \in \mc N$ (red circles) through its private action-reaction mapping, $f_i(\cdot)$, which may depend on the decision of any other agent in $\mathcal N$ (solid red lines), with the goal of predicting an outcome of the multi-agent interaction process.}
	\label{fig:scenario}
\end{figure}

Unlike the works above, we design an active learning scheme that allows an agnostic external entity to reproduce faithful approximations of action-reaction mappings, privately held by a population of agents, in order to predict a stationary profile of the underlying multi-agent interaction process exactly.

Specifically, we consider the scenario illustrated in Fig.~\ref{fig:scenario} and mathematically formalized in \S \ref{subsec:mathematical_formulation}, where such an external observer, endowed with a learning procedure $\mathscr{L}$, can get samples of the action-reaction mappings $f_i(\cdot)$ of each agent $i$, for all $i=1,\ldots,N$; in particular, the external observer iteratively proposes to agent $i$ a configuration $\hat{\bs x}_{-i}$ of possible other agents' decisions and gets back the reaction $x_i=f_i(\hat{\bs x}_{-i})$ of the agent to that configuration (see also the different tasks reported in Algorithm~\ref{alg:learning_embedded_BR}). 
The proposed scheme is of active-learning nature in that the queries $\hat{\bs x}_{-i}$ are generated based on the current surrogate models $\hat f_i(\cdot)$ of the agents' responses.

In our framework the low-level interaction among agents is not particularly relevant, since they are left free to possibly interact with each other. As such, we do not assume any particular communication pattern underlying the information exchange among the agents and topology of interactions. In fact, we only assume that the decision of some agent $i$ may be affected by the decision of any other agent within $\mathcal N$.
On the contrary, we focus on the external entity, which iteratively makes queries to estimate private action-reaction mappings with the ultimate goal of predicting a a stationary profile $\bs x^\star$ \emph{exactly}. To this end, the proposed learning mechanism allows the central entity to collect and exploit high-quality data that are informative \gls{wrt} such a stationary point prediction task. We summarize our contributions as follows:
\begin{enumerate}
	\item[i)] Formalize a novel problem setting involving an external entity that aims at predicting a possible outcome in multi-agent decision-making and control problems in which the decision policies of the agents are private  (\S \ref{sec:math_form}, \ref{sec:generalization});
	\item[ii)] Devise an active learning algorithm allowing the external entity to collect sensible data and update parametric estimates of the agents' action-reaction mappings (\S \ref{sec:learning_proc});
	\item[iii)] Prove sufficient conditions to assess the asymptotic properties of our active learning scheme so that, if convergence happens, it can only be towards a stationary action profile (\S \ref{sec:preliminary_results}, \ref{sec:learning_proc}). This fact entails two main implications:
	\begin{enumerate}
		\item Learning locally exact surrogates of the action-reaction mappings allows the external observer to succeed in its prediction task;
		\item Working with very general assumptions, the established conditions also serve as certificates for the existence of a stationary action profile.
	\end{enumerate}
	\item[iv)] A demonstration of the practical effectiveness of our methodology through extensive numerical simulations on the indirect control method for smart grids in \S \ref{subsec:motivating_example}, and typical multi-agent control and decision-making problems, including generalized Nash games and competitive linear feedback design problems (\S \ref{sec:applications_simulations}).
\end{enumerate}

The proofs of the technical results of the paper are all deferred to Appendix~\ref{sec:proofs}, while Appendix~\ref{sec:kalman_eq} reports mathematical derivations adopted in the implementation of our approach. The Python package \textsf{gnep-learn}, implementing the proposed active learning-based scheme, is publicly available at {\tt \hyperref{https://github.com/bemporad/gnep-learn}{}{}{https://github.com/bemporad/gnep-learn}}.

To the best of our knowledge, this work represents a first attempt integrating traditional machine learning paradigms within a smart query process, with the goal of predicting a possible outcome in multi-agent problems in which the decision policies of the agents are kept private.
In contrast with the cognate literature reviewed in \S \ref{subsec:literature}, we design a deterministic, active learning-based mechanism that exploits action-reaction samples to learn private action-reaction mappings, a setting that covers a wide variety of problems involving competitive agents with continuous (i.e., possibly infinite) decision sets.


\subsubsection*{Notation}
$\N$, $\R$ and $\R_{\geq 0}$ denote the set of natural, real, and nonnegative real numbers, respectively. $\N_0 \eqdef \N \cup \{0\}$, while $\bar\R\eqdef\R\cup\{+\infty\}$.
The transpose of a matrix $A \in \R^{n \times n}$ is $A^\top$, while
$A \succ 0$ ($\succcurlyeq 0$) denotes its positive (semi)definiteness. For a vector $v \in \R^n$ and a matrix $A \succ 0$, $\|v\|_2$ denotes the standard Euclidean norm, while $\|\cdot\|_A$ the $A$--induced norm $\|v\|_A \coloneqq \sqrt{v^\top A v}$.
The $n$-dimensional ball centred around $\bar x$ with radius $\theta > 0$ is $\mc{B}_\theta(\bar x) \eqdef \set{x \in \R^n}{\|x - \bar x\|_2 \leq \theta}$.
$I_{n}$, $\bsone_n$, and $\bs{0}_n$ denote the $n \times n$ identity matrix, the vector of all $1$, and $0$, respectively (we omit the dimension $n$ whenever clear). Given a set $\mc X \subseteq \R^n$, $\iota_\mc X:\R^n\to\bar\R$ denote the associated indicator function, i.e., $\iota_\mc X(x)=0$ if $x\in \mc X$, $\iota_\mc X(x)=+\infty$ otherwise. 
The operator $\col(\cdot)$ stacks its arguments in column vectors or matrices of compatible dimensions.
For example, given vectors $x_1,\dots,x_N$ with $x_i\in\mathbb{R}^{n_i}$ and $\mc N=\{1,\dots,N \}$, we denote $\bs{x} \eqdef (x_1 ^\top,\dots ,x_N^\top )^\top = \col((x_i)_{i\in\mc N}) \in \R^n$, $n \eqdef \sum_{i\in \mc N} n_i$, and $ \bs{x}_{-i} \eqdef \col(( x_j )_{j\in\mc N\setminus \{i\}})$. With a slight abuse of notation, we sometimes use also $\bs x = (x_i, \bs x_{-i})$.
The uniform distribution on the interval $[a,b]$ is denoted by $\mathcal U(a,b)$. 

\section{Learning problem}\label{sec:math_form}
We start by formalizing the learning problem addressed, which is successively motivated by two multi-agent control and decision-making problems that fit the framework we consider.
\subsection{Mathematical formulation}\label{subsec:mathematical_formulation}
We consider the scenario illustrated in Fig.~\ref{fig:scenario}. There, an external observer (or entity) is allowed to make queries and observe the reactions taken by a set of $N$ (possibly competitive) agents, indexed by $\mc N \coloneqq \{1, \ldots, N\}$, which mutually influence each other. Specifically, we assume that each agent controls a vector of variables $x_i \in \R^{n_i}$ to react, by means of a private \emph{action-reaction mapping} $f_i : \R^{n_{-i}} \to \R^{n_i}$, to the other agents' actions $\bs x_{-i} \in \R^{n_{-i}}$, $n_{-i} \eqdef \sum_{j \in \mc N \setminus \{i\}} n_j$.
Within the multi-agent interaction process, the whole population of agents has also to meet collective constraints, i.e., $\bs x \eqdef (x_i, \bs x_{-i}) \in \Omega \subseteq \R^n$, $n \eqdef n_i + n_{-i}$, which are
known to every agent and the external entity, and whose portion involving the $i$-th agent can be embedded directly within the action choice $f_i(\cdot)$. Specifically, for any given $\bs x_{-i}$, $f_i(\bs x_{-i})$ is so that $(f_i(\bs x_{-i}), \bs x_{-i}) \in \Omega$, for all $i \in \mc N$.
We assume that possible additional local constraints $x_i\in \mathcal{X}_i$, such as lower and upper bounds on $x_i$, are either private (i.e., unknown to the other agents and the external entity), or shared. In both cases, it always holds that the resulting values $f_i(\bs x_{-i})\in\mathcal{X}_i$. In the former case, this is a \emph{de facto} situation that the external observer and the other agents take cognizance of. In the latter case, we assume that $x_i\in \mathcal{X}_i$ is already embedded in the collective constraint $\bs x\in \Omega$. 

Being completely agnostic on the overall multi-agent interaction process and main quantities involved, we assume the underlying external entity endowed with some parametric learning procedure $\mathscr{L}$, which exploits non-private data (i.e., the decisions $f_i(\bs x_{-i})$) collected iteratively from the agents through a query process. Specifically, the main goal of the external entity is to learn, at least locally, faithful surrogates $\hat f_i$ (formally defined later in this section) of the unknown mappings $f_i(\cdot)$ in order to predict a desirable outcome of the multi-agent interaction process, here identified in accordance with the following definition:

\begin{definition}\textup{(Stationary action profile)}\label{def:stat_action}
	A collective action profile $\bs{x}^\star \in \Omega$ is \emph{stationary} if, for all $i \in \mc N$, $x^\star_i =  f_i(\bs x^\star_{-i})$.
	\hfill$\square$
\end{definition}

Given a stationary profile, which coincides with a fixed point of the action-reaction mappings, none of the agents has incentive to deviate from the action currently taken. We will discuss in \S \ref{sec:applications} (and \S \ref{subsec:composition} when introducing a generalization of the problem investigated) suitable multi-agent control and decision-making applications calling for a stationarity condition as the one in Definition~\ref{def:stat_action}.

We now state some conditions on the formalized problem that we assume will hold true throughout the paper:
\begin{standing}[\textup{Mappings and constraints}]\label{standing:standard}
	For all $i \in \mc N$, the single-valued mapping $f_i : \R^{n_{-i}} \to \R^{n_i}$ is continuous. In addition, the set $\Omega \subseteq \R^n$ of feasible collective actions is nonempty.
	\hfill$\square$
\end{standing}
While the non-emptiness of $\Omega$ is an obvious requirement for the problem to be solvable, continuity of $f_i$ is a technical, yet not very restrictive, condition needed to prove our results.

As mentioned above, we endow the external observer with a learning procedure $\mathscr{L}$ to produce faithful proxies of the action-reaction mappings $f_i(\cdot)$ executed by the agents. Specifically, we let $\hat f_i : \R^{n_{-i}} \times \R^{p_i} \to \R^{n_i}$ denote the mapping of agent $i$ as estimated by the external entity, which is parametrized by $\theta_i \in \R^{p_i}$ to be updated iteratively by integrating data obtained from the agents through a query process, as described in \S \ref{sec:learning_proc}. 

Since our technical analysis will only require $\hat f_i(\cdot,\theta_i)$ to approximate $f_i(\cdot)$ \emph{locally} around a possible stationary action, for simplicity, we parameterize $\hat f_i(\cdot,\theta_i)$ 
as the affine mapping
\begin{equation}\label{eq:linear-predictor}
	\hat f_i(\bs x_{-i}, \theta_i)=\Lambda_i\begin{bmatrix}\bs x_{-i}\\1\end{bmatrix},
\end{equation}
for a matrix of coefficients $\Lambda_i \in \R^{n_i \times (n_{-i}+1)}$ (in this case, $\theta_i$ is the vectorization of $\Lambda_i$, with $p_i=n_i(n_{-i}+1)$).
In \S \ref{subsec:convergence} we will show that adopting such simple surrogate mappings comes with no restrictions, in the sense that they actually guarantee the external entity to succeed in its prediction task. In addition, they bring significant simplifications in the procedure discussed in Algorithm~\ref{alg:learning_embedded_BR}, thus fully motivating their use.
The case involving a generic $\hat f_i(\cdot,\theta_i)$, instead, will be discussed in \S \ref{subsec:generic_surrogate}.


Finally, we remark that the conditions postulated in Standing Assumption~\ref{standing:standard} on the multi-agent interaction process are very general and, as a result, they do not even guarantee the existence of a stationary action profile in the sense of Definition~\ref{def:stat_action}. As a distinct feature of our approach, we will show that sufficient conditions can be established to assess the asymptotic properties of the proposed learning-based algorithm so that, if convergence happens, it can only be towards a stationary action profile, thus also providing certificates for the existence of the latter.

\subsection{Applications in multi-agent control and decision-making}\label{sec:applications}
\subsubsection{Learning equilibria in generalized Nash games}\label{subsec:GNEP}
A \gls{GNEP} \cite{facchinei2010generalized} typically involves a population of $N$ agents where each one of them aims at minimizing some cost function $J_i : \R^n \to \R$ that depends both on its own (locally constrained) decision $x_i \in \mc X_i \subseteq \R^{n_i}$, as well as on the decisions of all the other agents $\bs{x}_{-i}$. In addition, the selfish agents also compete for shared resources, and are thus required to satisfy coupling constraints, i.e., $(x_i, \bs x_{-i}) \in \Omega \subseteq \R^n$. The resulting game is hence defined by a collection of mutually coupled optimization problems:
\begin{equation}\label{eq:single_prob}
	\forall i \in \mc N : \left\{
	\begin{aligned}
		& \underset{x_{i} \in \mc X_i}{\textrm{min}} &&  J_i(x_{i}, \bs x_{-i}) \\
		& \textrm{ s.t. } && (x_i, \bs x_{-i}) \in \Omega .
	\end{aligned}
	\right.
\end{equation}

A key notion is represented here by the \gls{GNE}, which typically identifies a desirable outcome of the noncooperative multi-agent decision-making process. In particular, by introducing $f_i(\bs x_{-i}) = \textrm{argmin}_{y_i \in \mc X_i(\bs{x}_{-i})}  \; J_i(y_i, \bs{x}_{-i})$ as the so-called \gls{BR} mapping, single-valued under strict convexity of $x_i \mapsto J_i(x_i, \bs{x}_{-i})$, and $\mc X_i(\bs{x}_{-i}) \eqdef \set{x_i \in \mc X_i}{(x_i, \bs x_{-i}) \in \Omega}$, a \gls{GNE} $\bs x^\star$ amounts to a fixed point of the stack of the \gls{BR} mappings of the agents, i.e., $x^\star_i = f_i(\bs{x}_{-i}^\star)$ for all $i \in \mc N$.
Note that numerous standard problems in robust and model predictive control \cite{witsenhausen1968minimax,bemporad2003min,bacsar2008h} based on min-max optimization $\textrm{min}_{x_1} \ \textrm{max}_{x_2}~J(x_1,x_2)$
can be recast in the form in \eqref{eq:single_prob}, where $J_1(x_1,x_2)=J(x_1,x_2)$, $J_2(x_2,x_1)=-J(x_1,x_2)$, and the set $\Omega$ is the Cartesian product between the set of feasible inputs $x_1$
and the set of possible disturbances $x_2$.

In this way, a \gls{GNE} coincides with a stationary action profile as in Definition~\ref{def:stat_action}, and an external entity aims at reconstructing the \gls{BR} mappings of the agents to predict a possible outcome of the observed \gls{GNEP}, i.e., a \gls{GNE}. Also in this case, we note that Standing Assumption~\ref{standing:standard} is not sufficient to guarantee the existence of a \gls{GNE} profile.

\subsubsection{Multi-agent feedback controller synthesis}\label{subsec:dec_con}
Consider a competitive version of traditional decentralized control synthesis problems \cite{crusius1999sufficient,barcelli2010synthesis}, in which each of the $N$ agents wants to stabilize some prescribed output $y_i \in \R^{n_{y_i}}$ of a global dynamical system, a goal that may be shared with other agents, by manipulating a subvector $u_i\in\R^{n_{u_i}}$
of control inputs. The overall \gls{LTI} dynamics thus reads as:
\begin{equation}\label{eq:LTI}
	\left\{
	\begin{aligned}
		&z(k+1) = A z(k) + \sum_{i\in \mc N} B_i u_i(k),\\
		&y_i(k) = C_i z(k), \text{ for all } i \in \mc N,
	\end{aligned}
	\right.
\end{equation}
where $z \in \R^{n_z}$ denotes the full state vector, $A \in \R^{n_z \times n_z}$
and $B=[B_1\ \ldots\ B_N]\in \R^{n_z \times n_u}$, $n_u\eqdef\sum_{i\in \mc N}n_{u_i}$, the matrices of the state-update function, with $B_i \in \R^{n_z\times n_{u_i}}$, and $C_i \in \R^{n_{y_i} \times n_z}$ the $i$-th output matrix. 
We assume  each pair $(A,B_i)$ controllable and $(A,C_i)$ observable, for all $i \in \mc N$.
Note that while the input vector
$u = \col((u_i)_{i \in \mc N}) \in \R^{n_u}$
is partitioned in $N$ subvectors, the output vectors $y_i$ are combinations of the full state $z$, for example $y_i$ and $y_j$ might contain overlapping subvectors of $z$.
In case each agent adopts a linear feedback controller $u_i(z(k)) = \kappa_i z(k)$, the overall control design then reduces to synthesizing gains $\kappa_i \in \R^{n_{u_i}\times n_z}$. From the $i$-th agent's perspective, the state evolution in \eqref{eq:LTI} hence turns into:
$$
	z(k+1) = \underbrace{\left(A + \sum_{j\in\mc{N}\setminus\{i\}} B_j \kappa_j \right)}_{\reqdef A_i(\bs \kappa_{-i}) = A_i} z(k) + B_i \kappa_i z(k).
$$
In case each agent has full knowledge of the global model $(A, B)$ and is interested in minimizing a standard infinite-horizon cost as local performance index $J_i(\kappa_i,\bs \kappa_{-i})=\sum_{k=0}^{\infty}\left(y_i(k)^\top Q_i y_i(k) + u^\top_i(k)
R_iu_i(k)\right)$, $R_i \succ 0$ and $Q_i \succcurlyeq 0$, the $i$-th action-reaction mapping in this case reads as:
\ifTwoColumn
\else
	\begin{equation}\label{eq:action_reaction_LQR}
		f_i(\bs \kappa_{-i}) = \set{\kappa_i \in \R^{n_z}}{\begin{aligned}
				& \kappa^\top_i = \left(R_i+B_i^\top P_i B_i\right)^{-1} B_i^\top P_i A_i,\\
				&A_i^\top P_i A_i-A_i^\top P_i B_i \left(R_i+B_i^\top P_i B_i\right)^{-1} B_i^\top P_i A_i+\bar Q_i=0,\\
				&P_i \succcurlyeq 0
		\end{aligned}},
	\end{equation}
\fi
where $\bar Q_i \eqdef C_i^\top Q_i C_i$. Each agent thus aims at designing a standard \gls{LQR} trying to stabilize only the particular output of interest 
$y_i$ so that \eqref{eq:LTI} is steered to the origin. Note that whenever a solution to each \gls{ARE} in \eqref{eq:action_reaction_LQR} exists and is unique, every action-reaction mapping $f_i(\cdot)$ happens to be single-valued. 

However, since the evolution of each $y_i$ heavily depends on the linear gains $\kappa_j$, $j \in \mc N \setminus \{i\}$, the outputs $y_i$'s can be partially
overlapping, and the weights $Q_i$ are individually chosen,
a conflicting scenario among agents' decisions may be triggered.
The solution of the underlying multi-agent feedback controller synthesis thus calls for a collective action profile in the spirit of Definition~\ref{def:stat_action}. Note that, in case such a profile does exist, then it is necessarily stabilizing since \eqref{eq:action_reaction_LQR} requires that, for fixed $\bs \kappa_{-i}$, $(A_i(\bs \kappa_{-i}) + B_i \kappa_i)$ has eigenvalues strictly inside the unit circle for all $i \in \mc N$. Establishing the existence of a collective action profile a-priori, however, is less obvious.

\section{Preliminary results}\label{sec:preliminary_results}
We now provide general convergence results under certain assumptions, which 
will be then further specialized in \S \ref{subsec:convergence} to establish the properties of our active learning scheme. 

Let us then consider a generic function $r:\R^n\times\R^p\to\R$, evaluated as $r(x,\theta)$ 
with $x\in\R^n$ and $\theta\in\R^p$, and a feasible set $\mc X\subseteq\R^n$ satisfying the following conditions:
\begin{assumption}
	\label{ass:convexity-1}
	The set $\mc X$ is convex. Moreover, for all $\theta\in\R^p$, $x\mapsto r(x,\theta)$ is a convex function.
	\hfill$\square$
\end{assumption}

\begin{assumption}
	\label{ass:convexity-2}
	The set $\mc X$ is bounded and nonempty. Moreover, it holds that:
	\begin{enumerate}
		\item[\normaltext{(i)}] For all $x\in \mc X$, $\theta \mapsto r(x,\theta)$ is convex and differentiable;
		\item[\normaltext{(ii)}] For all $\theta\in\R^p$, $x\mapsto r(x,\theta)$ is continuous;
		\item[\normaltext{(iii)}] For all $\theta\in\R^p$, the vector $\partial r(x,\theta)/\partial \theta\in\R^p$ of partial derivatives is bounded \gls{wrt} $x$.\hfill$\square$
	\end{enumerate}
\end{assumption}

\begin{lemma}
	\label{lemma:convexity_fk}
	Let Assumption~\ref{ass:convexity-1} hold true. Then, given any $\theta\in\R^p$, the set
	$
		\mc M(\theta)\eqdef {\normaltext{\textrm{argmin}}}_{y\in \mc X}~r(y,\theta)
	$
	is convex. If also Assumption~\ref{ass:convexity-2} holds true, then
	the value function $r^\star:\R^p\to\R$, $r^\star(\theta)\eqdef\normaltext{\textrm{min}}_{y\in \mc X}~r(y,\theta)$, 
	is continuous. 
	\hfill$\square$
\end{lemma}


Before proceeding further, we recall some basic definitions from \cite{rockafellar2009variational} useful for the remainder of this section:
\begin{definition}[\textup{Lower semicontinuity, \cite[Def.~1.5]{rockafellar2009variational}}]
	\label{def:lsc}
	A function $h:\R^n\to\bar\R$ is
	\emph{lower semicontinuous} if, for all $\tilde x\in\R^n$,
	$\underset{x\rightarrow \tilde x}{\lim}\inf h(x)\eqdef\underset{\varepsilon\rightarrow 0}{\lim}\left(
	\underset{y\in \mc B_\varepsilon(\tilde x)}{\inf} h(y)\right)\geq h(\tilde x)$.
	\hfill$\square$
\end{definition}

\begin{definition}[\textup{Level-boundedness, \cite[Def.~1.16]{rockafellar2009variational}}]
	\label{def:level-bounded}
	A function $g:\R^n\times\R^p\to\bar\R$ is
	\emph{level-bounded in $x$ locally uniformly in $\theta$} if, 
	for all $\tilde\theta\in\R^p$ and $\alpha\in\R$,
	there exists a neighbourhood $\mc I = \mc I(\tilde\theta) \subseteq \R^p$ and 
	a bounded set $\mc S\subset\R^n$ such that $\set{x\in\R^n}{g(x,\theta)\leq\alpha} \subset \mc S$ for
	all $\theta\in \mc I$.
	\hfill$\square$
\end{definition}

\begin{lemma}
	\label{lemma:lsc_level-bounded}
	Let Assumption~\ref{ass:convexity-1} and \ref{ass:convexity-2} hold true. Moreover, let $\mc X$ be a closed set.
	Then, the function $\bar{r}:\R^n\times\R^p\to\bar\R$, defined as $\bar{r}(x,\theta) \eqdef r(x,\theta)+\iota_\mc X(x)$ is lower semicontinuous and level-bounded in $x$ locally uniformly in $\theta$.
	\hfill$\square$
\end{lemma}

We conclude this section with the following result establishing the convergence of the sequence of minimizers for $r$ with minimum norm, $\{x^k\}_{k\in\N}$. Specifically, given some $\theta^k\in\R^p$ the corresponding element $x^k$ is defined as follows:
\begin{equation}\label{eq:xk}
		x^k \eqdef \textrm{argmin}_{y \in \R^n} \set{\tfrac{1}{2}\|y\|_2^2}{y\in \mc M(\theta^k)}.
\end{equation}

\begin{lemma}
	\label{lemma:convergence}
	Let Assumption~\ref{ass:convexity-1} and \ref{ass:convexity-2} hold true,
	and assume that $\mc X$ is also closed. Let $\{\theta^k\}_{k\in\N}$ be a sequence
	so that $\lim_{k\rightarrow\infty}\theta^k=\tilde\theta$, and let $\mc M(\tilde\theta) = \{\tilde x\}$. Then, the sequence $\{x^k\}_{k\in\N}$ generated by \eqref{eq:xk} is feasible, i.e., $x^k \in \mc X$ for all $k\in\N$, and satisfies
	$
	\lim_{k\rightarrow\infty}x^k=\tilde x.
	$
	\hfill$\square$
\end{lemma}

The technical results just introduced will be instrumental to establish the asymptotic properties of the active learning scheme presented in the next section. Specifically, we will see in \S \ref{subsec:convergence} how the proposed method meets Assumption~\ref{ass:convexity-1} and \ref{ass:convexity-2} to exploit the convergence of the sequence of minimum-norm minimizers of a suitable collection of parametric programs.

\section{Active learning procedure and main results}\label{sec:learning_proc}
We now introduce, discuss and analyze the convergence property of our iterative scheme based on a active learning.

\subsection{Algorithm description}\label{subsec:algorithm}
\begin{algorithm}[!t]
	\caption{Active learning-based method}\label{alg:learning_embedded_BR}
	\smallskip
	\textbf{Initialization:} $\bs x^0 \in \Omega$, $\theta_i^0 \in \R^{p_i} \text{ for all } i \in \mc N$\\
	\smallskip
	\textbf{Iteration $(k \in \N_0)$:}
	\vspace{.1cm}
	\begin{itemize}
		\item[$\bullet$] $\theta_i^{k+1} \in  \underset{\xi_i \in \R^{p_i}}{\textrm{argmin}} \; \frac{1}{k} 
		\sum_{t = 1}^k \ell_i(x_i^t, \hat f_i(\hat{\bs x}^t_{-i}, \xi_i)), \forall i \in \mc N$
		\smallskip
		\item[$\bullet$] $\hat{\bs x}^{k+1}  = \underset{\bs y \in \R^n}{\textrm{argmin}}~\tfrac{1}{2}\|\bs y\|_2^2~\textrm{ s.t. }~\bs y \in \mc M(\theta^{k+1})$
		\smallskip
		\item[$\circ$] $\text{For all } i \in \mc N : x^{k+1}_i = f_i(\hat{\bs x}^{k+1}_{-i})$
	\end{itemize}
\end{algorithm}

The main steps of our active learning scheme are summarized in Algorithm~\ref{alg:learning_embedded_BR}\footnote{A Python package with its implementation is publicly available at {\tt \hyperref{https://github.com/bemporad/gnep-learn}{}{}{https://github.com/bemporad/gnep-learn}}.}, where the black-filled bullets refer to the tasks the external observer is required to perform, while the empty bullet to the one performed by the agents in $\mc N$. 

Thus, at the generic $k$-th iteration of Algorithm~\ref{alg:learning_embedded_BR}, the external entity updates the affine surrogate mappings $\hat f_i(\cdot,\theta_i^k)$ including the most recent samples according to the following rule:
\ifTwoColumn
\else
	\begin{equation}\label{eq:param_update}
		\theta_i^{k+1} \in \underset{\xi_i \in \R^{p_i}}{\textrm{argmin}} \; \frac{1}{k} \sum_{t = 1}^k \ell_i(x_i^t, \hat f_i(\hat{\bs x}^t_{-i}, \xi_i)) \reqdef \Theta_i(\left\{(x_i^t,\hat{\bs x}^t_{-i})\right\}_{t=1}^k) = \Theta_i^k, \text{ for all } i \in \mc N ,
	\end{equation} 
	where $\ell_i : \R^n \times \R^{p_i} \to \R$ is some \emph{loss function}, typically dictated by the function approximation type adopted and the problem at hand.
\fi
Referring to Algorithm~\ref{alg:learning_embedded_BR}, we preliminarily define, for all $i \in \mc N$, $\textrm{argmin}_{\xi_i \in \R^{p_i}} \; (1/k) \sum_{t = 1}^k \ell_i(x_i^t, \hat f_i(\hat{\bs x}^t_{-i}, \xi_i)) \reqdef \Theta_i(\left\{(x_i^t,\hat{\bs x}^t_{-i})\right\}_{t=1}^k) = \Theta_i^k$. By referring to Fig.~\ref{fig:scenario}, the vector $\hat{ \bs x}^t$ hence denotes the query point employed by the external entity at the $t$-th iteration to gather all the best responses $f_i(\hat{\bs x}^t_{-i})$.

Successively, by relying on the updated proxies for the agents' action-reaction mappings $\hat f_i$, the external entity chooses the next query point $\hat{\bs x}^{k+1}$ as the minimum norm strategy profile in the set $\mc M:\R^p\to2^{\Omega}$, $p \eqdef \sum_{i\in \mc N} p_i$, formally defined as:
\begin{equation}\label{eq:minimizers}
	\mc M(\theta^{k+1}) \eqdef 	\underset{\bs y \in \Omega}{\textrm{argmin}} \; \sum_{i\in \mc N} \left\|y_i - \hat f_i(\bs y_{-i}, \theta^{k+1}_i) \right\|^2_2,
\end{equation}
which contains all collective profiles that are the closest (according to the squared Euclidean norm, although different metrics could be used) to a fixed point of each $\hat f_i(\cdot, \theta_i^{k+1})$. This step is motivated by the notion of stationary action profile in Definition~\ref{def:stat_action}, which coincides with a fixed point of the stack of the agents' mappings, i.e., $x_i^\star = f_i(\bs x_{-i}^\star)$ for all $i\in \mc N$. Indeed, if $\hat f_i$ were exactly equal to $f_i$ and the minimum in~\eqref{eq:minimizers} were
zero, any $\bs x^\star\in\mc M(\theta^{k+1})$ would be a stationary action profile.
Note that, by referring to \eqref{eq:minimizers}, $\theta^{k+1}$ identifies the whole collection of parameters $\{\theta_i^{k+1}\}_{i\in\mc N}$ characterizing the surrogate mappings, which at every iteration represent the argument of the corresponding parameter-to-query mapping $\mc M(\cdot)$.
In the spirit of the results developed in \S \ref{sec:preliminary_results}, the cost function in \eqref{eq:minimizers} then takes on the same role as the generic function $r(\cdot)$ in that section, thus establishing an immediate connection between \eqref{eq:minimizers} and the point-to-set mapping $\mc M(\cdot)$ defined in Lemma~\ref{lemma:convexity_fk}.

\begin{figure}[t!]
	\centering
	\ifTwoColumn
\else
	\includegraphics[width=.6\textwidth]{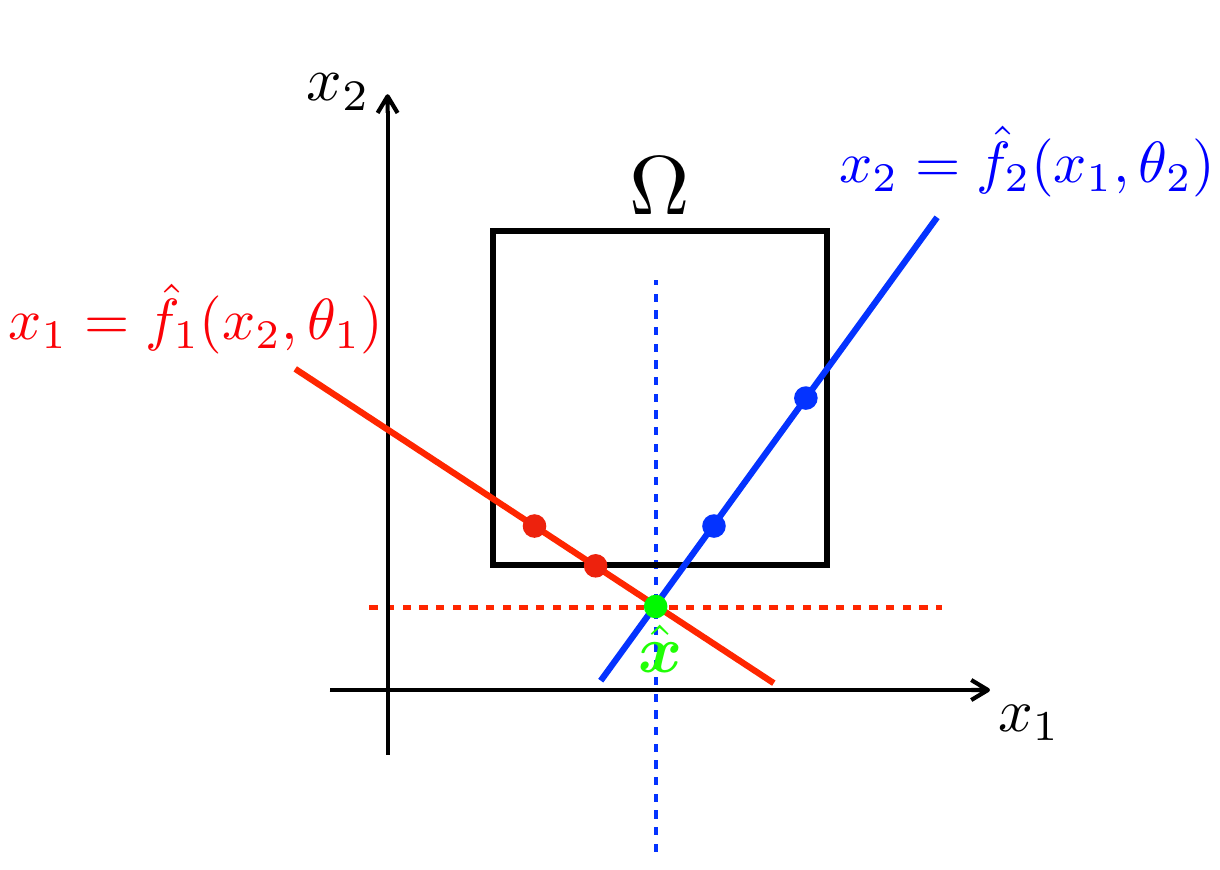}
	\fi
	\caption{With $N=2$, $n_1=n_2=1$, and affine $\hat f_i$'s constructed using feasible samples (red and blue dots), solving \eqref{eq:linear-system} would return the green point as a unique minimizer outside $\Omega$ (black box). While agent $2$ can still provide a feasible reaction to this infeasible query point (decision to be made along the dashed blue line), agent $1$ can not (decision along the dashed red line).}
	\label{fig:BR_example}
\end{figure}

By considering the minimization problem in \eqref{eq:minimizers}, we note that the special choice for the surrogate mappings in~\eqref{eq:linear-predictor} is particularly convenient, since \eqref{eq:minimizers} turns out to be a constrained \gls{LS} problem, which is convex whenever $\Omega$ (or  a conservative subset of it) is so.
To simplify notation, assume $n_i = 1$ for all $i \in \mc N$ so that $\Lambda_i= [\nu_i^\top \; c_i]$ in \eqref{eq:linear-predictor} (the extension to the case $n_i \ge 2$ follows readily). Then, the program in \eqref{eq:minimizers} becomes:
\begin{equation}
	\underset{\bs y \in\Omega}{\textrm{min}} \; \sum_{i\in \mc N} \left\|
	y_i-(\nu^{k+1}_i)^\top\bs y_{-i}-c^{k+1}_i
	\right\|^2_2.
	\label{eq:constrained-LS}
\end{equation}
Moreover, in the absence of constraints ($\Omega=\R^n$), we can further characterize the stationary action profile we wish to compute. In fact, problem~\eqref{eq:minimizers} can be solved by imposing $y_i - \hat f_i(\bs y_{-i}, \theta_i) =  y_i - \nu^\top_i \bs y_{-i} - c_i = 0$ for all $i \in \mc N$, therefore getting the following system of $N$ equations with $N$ unknowns:
\begin{equation}
	\begin{bmatrix}
		1 & & -\nu_1^\top \\
		\vdots & \ddots & \vdots\\
		-\nu_N^\top & & 1
	\end{bmatrix} \bs y = \begin{bmatrix}
		c_1\\
		\vdots\\
		c_N
	\end{bmatrix},
	\label{eq:linear-system}
\end{equation} 
where the superscript $^{k+1}$ has been omitted for simplicity.
In particular, if~\eqref{eq:linear-system} is solvable, every solution lies on a subspace 
$\mc M(\theta^{k+1})$ of dimension smaller than $N$ (a single point when it is unique),
and the vector $\hat{\bs x}^{k+1}$ corresponds to the minimum-norm solution of~\eqref{eq:linear-system}.
When $\Omega\subset\R^n$, one might still be tempted to solve~\eqref{eq:linear-system}
to get the next query point $\hat{\bs x}^{k+1}$. However, further queries may return infeasible reactions by some of the agents, as shown in the pictorial representation in Fig.~\ref{fig:BR_example}.

Once obtained the minimum norm vector $\hat{\bs x}^{k+1}$ in \eqref{eq:minimizers}, this serves as a query profile to collect the next agents' reaction data, $x^{k+1}_i = f_i(\hat{\bs x}^{k+1}_{-i})$. Within the final step of the procedure, indeed, $\hat{\bs x}^{k+1}$ is broadcasted to the agents, which on their side react to $\hat{\bs x}^{k+1}_{-i}$ by computing $f_i(\cdot)$, and finally return the results to the external entity. Therefore, each $x^{k+1}_i $ enriches the training dataset used in~\eqref{eq:param_update} to improve the affine model $\hat f_i(\cdot, \theta_i^{k+1})$.

\subsection{Initialization}
\label{sec:initialization}
As commonly done in most active learning algorithms~\cite{Set12,KG20,Bem23},
the execution of Algorithm~\ref{alg:learning_embedded_BR} is preceded by a (passive) random data collection phase to make an initial estimate of the affine surrogates $\hat f_i$. During this initialization procedure, the external entity draws iteratively a random feasible point $\bs x^k\in\Omega$ as follows. First, it forms
a random vector $\bs x_r^k\in\R^n$ by extracting
its components $x_{r_i}^k$ from distribution $\mc U(m^-_i,m^+_i)$
where $m^-\leq \bs x\leq m^+$ define a range of plausible values of $\bs x$. Then,
the following constrained \gls{LS} problem 
\begin{equation}
	\bs x^k=\underset{\bs y\in\Omega}{\textrm{argmin}} \; \|\bs y- \bs x_r^k\|_2^2, 
\label{eq:least-squares-init}
\end{equation}
is solved for $k=1,\ldots,K_{\textrm{in}}$ to define $\bs x^k$ and pass it in parallel to each agent, which in turn responds through the private action-reaction mapping $f_i$. The resulting
actions are collected by the external entity to update its estimates $\theta_i^k$. 
The remaining $K-K_{\textrm{in}}$ iterations of Algorithm~\ref{alg:learning_embedded_BR} are then executed,
where $K$ is the total number of queries performed to the agents. Alternatively, more sophisticated initialization strategies are possible, such as Latin hypercube sampling \cite{stein1987large,viana2013things} as in, e.g.,~\cite{Bem23}, however, significant differences are not expected in our framework as we only learn {\it linear} proxies
rather than global surrogates of the agents' responses. Nevertheless, given its key role we will discuss in detail the random initialization procedure within each numerical example reported in \S \ref{sec:applications_simulations}.

\subsection{Complexity analysis}
\label{sec:complexity_analysis}
We analyze the complexity of Algorithm~\ref{alg:learning_embedded_BR} in the affine setting~\eqref{eq:linear-predictor} along with the \gls{MSE} loss
\begin{equation}
\ell_i(x_i^t, \hat f_i(\hat{\bs x}^t_{-i}, \xi_i)) =\|x_i^t-\hat f_i(\hat{\bs x}^t_{-i}, \xi_i)\|_2^2,
\label{eq:MSE-loss}
\end{equation}
which make \eqref{eq:param_update} a linear \gls{LS} problem
and the acquisition problem~\eqref{eq:minimizers} either a linear system as in~\eqref{eq:linear-system}
($\Omega=\R^n$), or a constrained linear \gls{LS} as in~\eqref{eq:constrained-LS}
($\Omega\subset\R^n$ is a polyhedron), both having $N$ unknowns. Hence, the complexity of solving~\eqref{eq:minimizers} or~\eqref{eq:least-squares-init} is either the complexity of solving a linear system of $N$ equations or, in the constrained case, a linear \gls{LS} problem with $N$ residuals and as many inequalities as in the hyperplane representation of $\Omega$, multiplied by the number $K$ of iterations.
The memory size required to store the model parameters and form problem~\eqref{eq:minimizers} is therefore $\sum_{i=1}^n n_i(n_{-i}+1)$.

Due to the monotonically increasing number of training data $k$, we consider simple recursive methods to update the estimate $\theta_i^{k+1}$ in~\eqref{eq:param_update}, as described for instance in~\cite{ljung1983theory,kailath2000linear},
such as recursive \gls{LS}~\cite{AG93}. To control the learning rate of the algorithm, we adopt the following linear Kalman filter iterations~\cite{kalman1960} tailored for $n_i=1$ (see Appendix~\ref{sec:kalman_eq} for the derivation):
\begin{subequations}
	\begin{align}
		&\phi_i^k=\begin{bmatrix}\hat{\bs x}^k_{-i}\\1\end{bmatrix},\ a_i^k=P_i^k\phi_i^k\\
		&P^{k+\nicefrac{1}{2}}_i=P_i^k-\frac{a_i^k(a_i^k)^\top}{1+(\phi_i^k)^\top a_i^k}\\ 
		&\theta_i^{k+1} = \theta_i^{k} + P_i^{k+\nicefrac{1}{2}}\phi_i^k(x_i^k-(\phi_i^k)^\top\theta_i^k)\\
		& P_i^{k+1}=P_i^{k+\nicefrac{1}{2}}+\beta I
	\end{align}
	\label{eq:KF}%
\end{subequations}
For $n_i>1$, a bank of Kalman filters as in~\eqref{eq:KF} is run in parallel
for each of the $n_i$ components of $\hat f_i$. 
The complexity of solving~\eqref{eq:param_update} to update the parameter vector $\theta$ is hence $K\sum_{i=1}^n n_iO(n_{-i}^2)$ and requires
$\sum_{i=1}^n n_i(n_{-i}+1)(n_{-i}+2)/2$ additional numbers to store
the symmetric covariance matrices.

In~\eqref{eq:KF} $\beta\geq 0$ defines the learning rate of the filter, and the initial value of the $N$-by-$N$ covariance matrix
$P_i^0=\alpha I$ the amount of $\ell_2$-regularization on $\theta_i$ (cf.~\cite[Eq.~(12)]{Bem23b}). In particular, the larger the $\beta$, the faster is the convergence of the linear
estimator (we obtain standard recursive \gls{LS} updates for $\beta=0$~\cite{ljung1983theory,kailath2000linear}), while the smaller the $\alpha$, the larger is the $\ell_2$-regularization. 
We will adopt the above setting in all the numerical simulations reported in \S \ref{sec:applications_simulations}. 

\subsection{Asymptotic properties}\label{subsec:convergence}
Before characterizing the asymptotic properties of the active learning procedure in Algorithm~\ref{alg:learning_embedded_BR}, we first postulate some assumptions on the learning procedure $\mathscr{L}$ the external observer is endowed with and then establish some results
that will be instrumental for the main statement of the paper.
 
\begin{standing}[\textup{Training loss}]\label{standing:learning_procedure}
	For all $i \in \mc N$, the loss function $\ell_i$ is continuous in both arguments and, for all $(x_i, \bs x_{-i}, \theta_i) \in \Omega \times \R^{p_i}$, $0 \le \ell_i(x_i, \hat f_i(\bs x_{-i}, \theta_i)) < \infty$, with $\ell_i(x_i, \hat f_i(\bs x_{-i}, \theta_i)) = 0$ if and only if $x_i = \hat f_i(\bs x_{-i}, \theta_i)$.
	\hfill$\square$
\end{standing}
The conditions in Standing Assumption~\ref{standing:learning_procedure} allow us to characterize the choice of the loss functions adopted in \eqref{eq:param_update}. 

Our technical analysis will be based on the possibility of matching \emph{pointwise} (and, therefore, not necessarily globally nor even locally) the action-reaction mapping $f_i$ of each agent, which is a key property possessed by the affine surrogates in \eqref{eq:linear-predictor}, which can hence be used with no restrictions.
Specifically, for all $i \in \mc N$ and for all $(x_i, \bs x_{-i}) \in \Omega$, there exists a set $\mc A_i = \mc A_i(x_i, \bs x_{-i}) \subseteq \R^{p_i}$ such that $\ell_i(x_i, \hat f_i(\bs x_{-i}, \tilde\theta_i)) = 0$, for all $\tilde\theta_i \in \mc A_i$. While this condition is intrinsic for affine surrogates, it shall therefore be assumed to hold true in case of generic proxies $\hat f_i(\cdot,\theta_i)$ within the discussion in \S \ref{subsec:generic_surrogate}. 
Moreover, in view of the specific structure \eqref{eq:linear-predictor} we consider, the affine proxies are also continuous with respect to their second argument, i.e., $\hat f_i(\bs x_{-i}, \cdot)$ is continuous for all $\bs x_{-i} \in \R^{n_{-i}}$.

The crucial result stated next requires the external entity to solve \eqref{eq:param_update} at the global minimum at every iteration. In case there exist multiple (global) minimizers we then tacitly assume further the external observer being endowed with a tailored \emph{tie-break rule}, i.e., some single-valued mapping $\mc T : 2^{\R^{p_i}}\to\R^{p_i}$ used consistently across iterations, such as picking up the solution $\theta^k_i$ with minimum norm at every $k \in \N$, to single-out one of those minimizers, for all $i \in \mc N$. With this tie-break rule $\mc T$ in place, which takes a set of vectors in $\R^{p_i}$ as argument and returns one element from it according to some criterion, the first step in Algorithm~\ref{alg:learning_embedded_BR} performed by the external observer turns into an equality rather than an inclusion. 
\begin{lemma}\label{lemma:local_exactness}
	For all $i \in \mc N$, let $\{(x_i^t,\hat{\bs x}^t_{-i})\}_{t = 1}^k$ be a collection of samples to be employed in the update rule for the parameter $\theta_i$ in \eqref{eq:param_update}, where each $(x_i^t,\hat{\bs x}^t_{-i}) \in \Omega$. Assume that, for every $k \in \N$, \eqref{eq:param_update} is solved at the global minimum and a tie-break rule $\mc T$ is applied to single-out a minimizer. If $\lim_{k \to \infty} x_i^k = \bar{x}_i$, $\lim_{k \to \infty} \hat{\bs x}^k_{-i} = \bar{\bs x}_{-i}$ so that $(\bar{x}_i, \bar{\bs x}_{-i})\in\Omega$, and $\lim_{k \to \infty} \theta_i^k = \bar{\theta}_i$, then $\bar{\theta}_i \in \mc A_i(\bar{x}_i, \bar{\bs x}_{-i})$, for all $i \in \mc N$.
	\hfill$\square$
\end{lemma}
Lemma~\ref{lemma:local_exactness} shows a sort of ``consistency property'' of the affine surrogates $\hat f_i$, meaning that if all the ingredients involved in Algorithm~\ref{alg:learning_embedded_BR} happen to converge, then the pointwise approximation of $f_i$ shall be exact
at ${\bs x}_{-i}$, i.e., $\hat f_i({\bs x}_{-i},\tilde\theta_i)=f_i({\bs x}_{-i})$. 
\vspace{-.4cm}
	\begin{standing}[\textup{Feasible set $\Omega$}]\label{standing:surrogate_set}
		The set $\Omega$ is a nonempty polytope.
		\hfill$\square$
	\end{standing}
	We now characterize the properties of the sequence of query points $\{\hat{\bs x}^k\}_{k\in\N}$ produced by the central entity in the second step of Algorithm~\ref{alg:learning_embedded_BR} by computing the optimal solution to:
	\begin{equation}\label{eq:bilevel}
		\hat{\bs x}^k \eqdef \textrm{argmin}_{\bs y \in \R^n} \set{\tfrac{1}{2}\|\bs y\|_2^2}{\bs y \in \mc M(\theta^k)},
	\end{equation}
	for all $k \in \N$, thus mirroring the strategy to generate $x^k$ in \eqref{eq:xk}.
\begin{proposition}
	\label{prop:f}
	Let $\{\theta^k\}_{k\in\N}$ be a sequence
	so that $\lim_{k \to \infty} \theta_i^k = \tilde{\theta}_i$ for all $i \in \mc N$, and let $\mc M(\tilde\theta)=\{\tilde{\bs x}\}$. Then the
	sequence $\{\hat{\bs x}^k\}_{k\in\N}$ generated by \eqref{eq:bilevel} is feasible and satisfies
	$
	\lim_{k\rightarrow\infty}\hat{\bs x}^k=\tilde{\bs x}.
	$
	\hfill$\square$
\end{proposition}

 
We are now ready to state the asymptotic properties of the active learning methodology described in Algorithm~\ref{alg:learning_embedded_BR}:
\begin{theorem}\label{th:convergence}
	Let $\{\theta^k\}_{k\in\N}$ be a sequence
	so that $\lim_{k \to \infty} \theta_i^k = \tilde{\theta}_i$ for all $i \in \mc N$, and let $\mc M(\tilde \theta) = \{\tilde{\bs x}\}$.
	For every $k \in \N$, let \eqref{eq:param_update} be solved at the global minimum and a tie-break rule $\mc T$ applied to single-out a minimizer. Then, $\lim_{k \to \infty} \|\hat{\bs x}^k - \bs x^k\|_2 = 0$, and the sequence $\{\bs x^k\}_{k \in \N}$ and $\{\hat{\bs x}^k\}_{k \in \N}$ generated by Algorithm~\ref{alg:learning_embedded_BR} converge to the same stationary action profile.~\hfill$\square$
\end{theorem}
	As a consequence of Theorem~\ref{th:convergence}, 
	in case the convergence of the parametric estimates happens, it can only be towards the true (at least pointwise) values. This fact, together with $\lim_{k \to \infty} \|\hat{\bs x}^k - \bs x^k\|_2 = 0$, yields $\sum_{i\in \mc N} \|\tilde x_i - \hat f_i(\tilde{ \bs x}_{-i}, \tilde \theta_i)\|^2_2 = \sum_{i\in \mc N} \|\tilde x_i - f_i(\tilde{ \bs x}_{-i})\|^2_2 = 0 = \sum_{i\in \mc N} \|f_i(\bar{\bs{x}}_{-i}) - \bar x_i\|^2_2=0$, i.e., $\|f_i(\bar{\bs{x}}_{-i}) - \bar x_i\|_2=0$, i.e., the external entity has in its hands both a stationary action profile $\tilde{ \bs x}$, and pointwise-exact surrogates of the action-reaction mappings $\hat f_i(\tilde{ \bs x}_{-i}, \tilde \theta_i) = f_i(\tilde{ \bs x}_{-i})$. 
	
	Note that the assumptions postulated in \S \ref{subsec:mathematical_formulation} are not sufficient to guarantee the existence of a vector
	$\bs x^\star$ such that $\left\|x^\star_i - f_i(\bs x^\star_{-i}) \right\|_2 = 0$ for all $i \in \mc N$.
	The following corollary offers a certificate for the existence of at least a stationary action profile, and it is an immediate consequence of Theorem~\ref{th:convergence}.
\begin{corollary}\label{cor:existence}
	Under the conditions stated in Theorem~\ref{th:convergence}, the multi-agent interaction process at hand admits a stationary action profile $\bs x^\star$  in the sense of Definition~\ref{def:stat_action}.
	\hfill$\square$
\end{corollary}

\section{Technical discussion and possible extensions}\label{sec:generalization}
Next, we elaborate around the conditions granting the asymptotic properties in Theorem~\ref{th:convergence}, along with possible extensions to a more general framework or surrogate mappings.

\subsection{Discussion on Theorem~\ref{th:convergence}}\label{subsec:discussion}
Assuming that $\lim_{k \to \infty} \theta_i^k = \tilde{\theta}_i$ for all $i \in \mc N$, along with the existence of a single minimizer contained in $\mc M(\tilde\theta)$ \eqref{eq:minimizers}, appear rather strong requirements to achieve the convergence of the proposed active learning scheme that, to the best of our knowledge, can not be guaranteed beforehand, not even by imposing further structure on the learning procedure $\mathscr{L}$ or the multi-agent problem at hand. Both can indeed be verified only a posteriori or in practice after a large enough number of iterations of Algorithm~\ref{alg:learning_embedded_BR}. As a further motivation for adopting the affine surrogates in \eqref{eq:linear-predictor}, note that checking whether $\mc M(\tilde\theta)$ is a singleton translates into verifying the positive definiteness of the Hessian matrix associated to the cost function in \eqref{eq:constrained-LS}.

We note that, however, the technical conditions assumed in Standing Assumption~\ref{standing:standard} are so general that they do not even seem to be sufficient to establish the existence of a stationary action profile for the underlying problem. This should be put together with the fact that the external observer: i) has very little knowledge of such a process, ii) can be endowed with any learning procedure $\mathscr{L}$ satisfying Standing Assumption~\ref{standing:learning_procedure}, and iii) could succeed in its prediction task even with a simple, i.e., affine, parametrization for the surrogate mappings $\hat f_i$.

We point to \S \ref{sec:applications_simulations} for many numerical results in which the a-posteriori verification of the convergence of the parametric estimates yielding a single minimizer in $\mc M(\cdot)$ succeeds, thus demonstrating the practical effectiveness of our methodology.

\subsection{Learning composition of action-reaction mappings}\label{subsec:composition}
Consider the case in which the decision variable $x_i$ does not follow directly as a result of some reaction process through $f_i(\cdot)$, but it coincides with a function of some other inner variable, say $w_i \in \mc W_i\subseteq\R^{m_i}$ (in this new framework, $f_i$ should be suitably redefined taking $\R^{m_i}$ as codomain). By introducing $g_i : \R^{m_i} \to \R^{n_i}$ as an additional, private mapping we have that $x_i$ is obtained through the composition $(g_i \circo f_i)(\bs x_{-i})$, i.e.,
$$
	x_i = g_i(w_i) = g_i(f_i(\bs x_{-i})).
$$

This different perspective
 makes the reaction process to $\bs x_{-i}$ indirect.
\ifTwoColumn
\else
As a motivating example, consider a population of agents obeying to the following dynamics:
\begin{equation}\label{eq:dynamics}
	\forall i \in \mc N : \left\{
	\begin{aligned}
		& z_i(k+1) = d_i(z_i(k), u_i(k)) + s_i(\bs u_{-i}(k), \bs v_{-i}(k)),\\
		& v_i(k) = t_i(z_i(k),u_i(k)),
	\end{aligned}
	\right.
\end{equation}
\fi
where $z_i \in \R^{n_{z_i}}$, $u_i \in \R^{n_{u_i}}$, and $v_i \in \R^{n_{v_i}}$ denote the state vector, the control input, and the controlled output of each agent, respectively, whose evolution (except for $u_i$, which has to be designed) is determined by mappings $d_i : \R^{n_{z_i}} \times \R^{n_{u_i}} \to \R^{n_{z_i}}$ and $t_i : \R^{n_{z_i}} \times \R^{n_{u_i}} \to \R^{n_{v_i}}$. The behaviour of the state variables, however, is also affected by some function $s_i : \R^{n_{u_{-i}}} \times \R^{n_{v_{-i}}}$ of the other agents' control inputs and outputs. 
Planning an optimal control strategy over some horizon of length $T\ge1$ traditionally requires each agent to find a solution to the following optimal control problem:
\begin{equation}\label{eq:OCP}
	\forall i \in \mc N : \left\{
	\begin{aligned}
		&\underset{w_i}{\textrm{min}} &&  \Psi_i(z_i(T))+\sum_{k=0}^{T-1} \, \Phi_i(x_i(k), \bs x_{-i}(k),k) \\
		& \textrm{ s.t. } &&z_i(0)=z_{i,0}, \ \eqref{eq:dynamics},~\forall k = 1,\ldots,T-1,\\
		&&& x_i(k) \in \mc \mc X_i,~\forall k = 0,\ldots,T-1,
	\end{aligned}
	\right.
\end{equation}
where $z_{i,0} \in \R^{n_{z_i}}$ is the measured initial state, 
$w_i(k)=\col(u_i(k),z_i(k+1))$, $w_i \eqdef \col((w_i(k))_{k=0}^{T-1}) \in \R^{n_{w_i}}$, $n_{w_i} \eqdef T(n_{u_i}+n_{z_i})$, 
$x_i(k) \eqdef \col(u_i(k), v_i(k)) \in \R^{n_{u_i}+n_{v_i}}$, with $x_i\eqdef g(w_i)=\col((x_i(k))_{k=0}^{T-1}) \in \R^{n_i}$, $n_i \eqdef T(n_{u_i}+n_{v_i})$, 
$g:\R^{n_{w_i}}\to\R^{n_i}$,
$\mc \mc X_i \subseteq \R^{n_{u_i}}\times \R^{n_{v_i}}$ denotes the input-output constraint set, $\Phi_i : \R^{n_{u_i}+n_{v_i}}\times\R^{n_{x_{-i}}}\times\N \to \R$ is the $i$-th stage cost that depends on $k$ to include a possible reference output signal to track, and 
$\Psi_i : \R^{n_{z_i}}\to\R$ is a terminal cost.

The exogenous term $s_i$ in \eqref{eq:dynamics} thus makes \eqref{eq:OCP} a collection of optimization problems with coupling terms dependent on $\bs x_{-i}$ both in the cost and constraints, whose overall solution, if any, naturally calls for a collective action profile in the spirit of Definition~\ref{def:stat_action}, which in this generalized setting reads as:
$$\bs{x}^\star \in \Omega \text{ s.t., for all } i\in\mc N,~x^\star_i = g_i(f_i(\bs x^\star_{-i})).$$

The agnostic external entity will therefore collect data by querying the agents with tailored opponents' input/output profiles, and then update proxies for each composition $(g_i \circo f_i)$ to predict the optimal control actions adopted by the set of agents over the $T$-steps long horizon.

From a merely technical perspective, to let our machinery work also in this more general setting, the conditions on $f_i(\cdot)$ in Standing Assumption~\ref{standing:standard} shall be replaced with the following:
\begin{enumerate}
	\item[i)] Each mapping $(g_i \circo f_i)(\cdot)$ is continuous and single-valued;
	\item[ii)] For all $\bs x_{-i} \in \R^{n_{-i}}$, $f_i(\bs x_{-i})$ produces some $w_i\in \mc W_i$ so that $(g_i(w_i), \bs x_{-i}) \in \Omega$.
\end{enumerate}
Then, nothing prevents the external entity from adopting the same affine surrogate mappings \eqref{eq:linear-predictor} to estimate each composition $(g_i \circo f_i)(\cdot)$, whereas the final step of the active learning scheme in Algorithm~\ref{alg:learning_embedded_BR} will require to simply collect $x^{k+1}_i = g_i(f_i(\hat{\bs x}_{-i}^{k+1}))$ for all $i \in \mc N$. The technical analysis on the asymptotic properties of the variant of Algorithm~\ref{alg:learning_embedded_BR} just discussed then mimics the one carried out in \S \ref{subsec:convergence}.

\subsection{Generic surrogate mappings $\hat f_i$}\label{subsec:generic_surrogate}
Adopting the affine parametrization in \eqref{eq:linear-predictor} for the action-reaction proxies has been shown beneficial to satisfy the assumptions postulated so far, as well as to simplify the tasks in Algorithm~\ref{alg:learning_embedded_BR} the central entity has to perform. More importantly, in view of the analysis in \S \ref{subsec:convergence}, such simple surrogates can be used with no restrictions, since they actually guarantee the external entity to succeed in its prediction task.

In case, however, one is interested in more flexible surrogate mappings $\hat f_i$, some further conditions have to be imposed. Specifically, by mirroring the discussion in \S \ref{subsec:convergence} tailored for affine proxies, one shall necessarily integrate Standing Assumption~\ref{standing:learning_procedure} with a requirement on the continuity of the mapping $\theta_i \mapsto \hat f_i(\bs x_{-i},\theta_i)$ for all $\bs x_{-i} \in \R^{n_{-i}}$. Nevertheless, the proxies $\hat f_i$ shall also be able to match at least pointwise the action-reaction mapping $f_i$ of each agent, thus requiring that for all $(x_i, \bs x_{-i}) \in \Omega$, there exists a set $\mc A_i = \mc A_i(x_i, \bs x_{-i}) \subseteq \R^{p_i}$ such that $\ell_i(x_i, \hat f_i(\bs x_{-i}, \tilde\theta_i)) = 0$, for all $\tilde\theta_i \in \mc A_i$. These two conditions together are key to prove the consistency property in Lemma~\ref{lemma:local_exactness} also with generic surrogate mappings.
In addition, to rely on Proposition~\ref{prop:f} one should also ensure that the chosen proxies $\hat f_i$ allow meeting Assumption~\ref{ass:convexity-1} and Assumption~\ref{ass:convexity-2}, which characterize the second task performed by the external entity in Algorithm~\ref{alg:learning_embedded_BR}, so that the arguments made in the proof of Theorem~\ref{alg:learning_embedded_BR} could be applied also to this more generic case.

A main drawback of using nonlinear surrogates is that the cost function in problem~\eqref{eq:minimizers} may easily become nonconvex, unless combined with suitable loss functions. For example, a convex program equivalent to \eqref{eq:minimizers} might be obtained by adopting softmax surrogate mappings and cross-entropy losses, in place of the Euclidean norm, to model categorical decision variables. This is however a topic of current investigation.

\section{Numerical results}\label{sec:applications_simulations}
We now test the practical effectiveness of the active learning scheme in Algorithm~\ref{alg:learning_embedded_BR} on several numerical instances of the multi-agent control and decision-making problems described in \S \ref{subsec:motivating_example}, \S \ref{sec:math_form}, along with comparing its performance with the state-of-the-art centralized method for \gls{GNE} computation in \cite{hespanha2022tenscalc}. We adopt the affine setting~\eqref{eq:linear-predictor} with \gls{MSE} loss~\eqref{eq:MSE-loss} detailed in \S \ref{sec:complexity_analysis}. All simulations are run in MATLAB on a laptop with an Apple M2 chip featuring an 8-core CPU and 16 GB RAM. Some of the examples analyzed in this section are also included in the Python package \textsf{gnep-learn}, publicly available at {\tt \hyperref{https://github.com/bemporad/gnep-learn}{}{}{https://github.com/bemporad/gnep-learn}}.

\begin{table}[tb]
	\caption{Charging \glspl{EV} -- Simulation parameters}
	\label{tab:sim_val}
	\centering
	\begin{tabular}{lll}
		\toprule
		Parameters  &  Description   & Value \\
		\midrule
		$T$ & Time interval & $14$\\
		$N$ & Number of \glspl{EV}  & $10$\\
		$q_i$& Degradation cost -- quadratic term & $\sim\mc U(0.006,0.01)$\\
		$c_i$& Degradation cost -- affine term & $\sim\mc U(0.055,0.095)^T$\\
		$d$& Normalized inflexibility demand & from \cite[Fig.~1]{ma2011decentralized}\\
		$\gamma_i$ & Local charging requirement &$\sim\mc U(1.2,1.8)$\\
		$\bar c_i$ & Upper bound - power injection &$0.25$\\
		$\bar c$ & Grid capacity &$0.2$\\
		$K$ & Number of iterations (Alg.~\ref{alg:learning_embedded_BR}) &$200$\\
		$K_\textrm{in}$ & \% of $K$ - initialization of Alg.~\ref{alg:learning_embedded_BR} & $20$\\
		$\beta$ & Learning rate (Kalman filter) &$1$\\
		\bottomrule
	\end{tabular}
\end{table}

\begin{figure}[!t]
	\centering
	\ifTwoColumn
\else
	\includegraphics[width=.6\columnwidth]{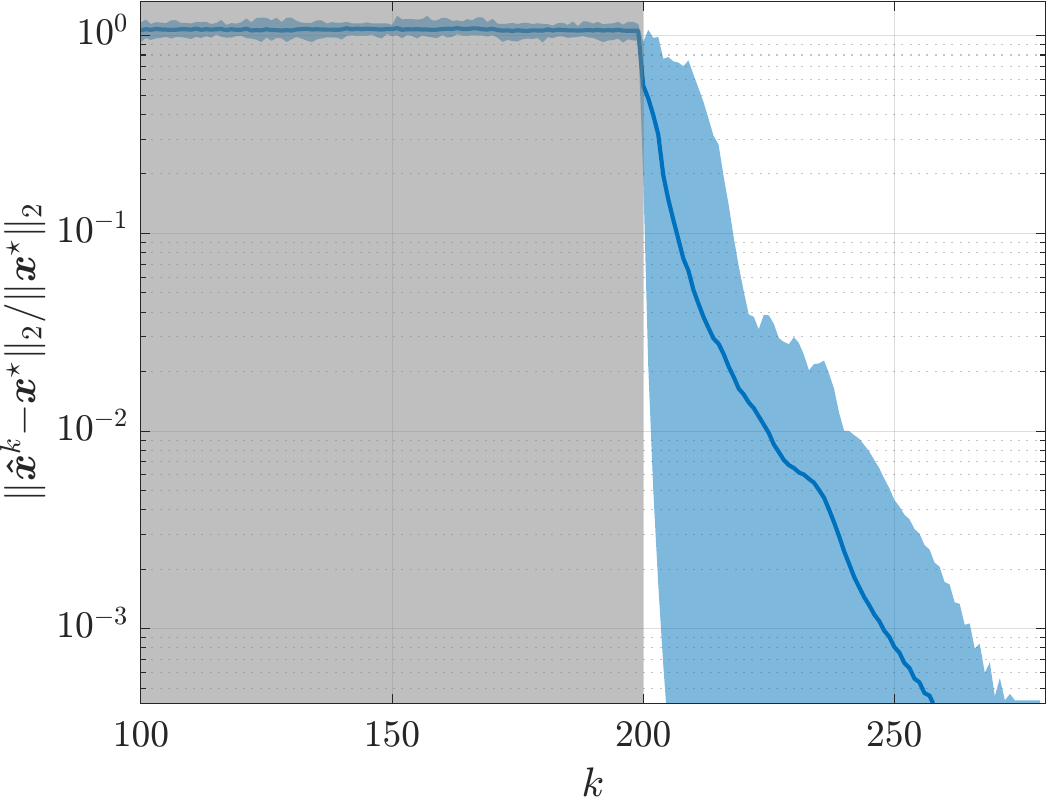}
	\fi
	\caption{Sequence $\{\hat{\bs x}^k\}_{k\in\N}$, averaged over $25$ numerical instances (solid blue line). The shaded black region corresponds to the passive random data collection phase, which is reported for the interval $k\in[100,200]$ only for illustrative purposes.}
	\label{fig:ev_convergence}
\end{figure}

\begin{figure}[!t]
	\centering
	\ifTwoColumn
\else
	\includegraphics[width=.6\columnwidth]{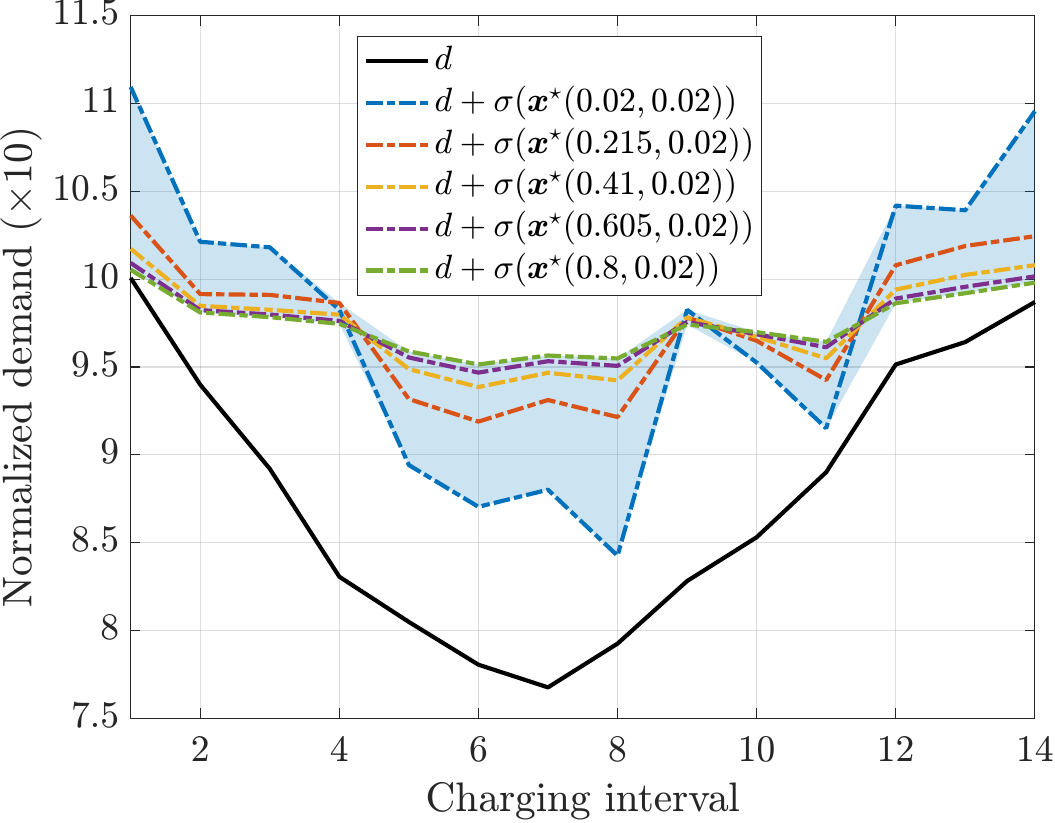}
	\fi
	\caption{Sum between the normalized average inflexible demand $d$ and a sample of five charging strategies $(1/N) \sum_{i\in \mc N} x_i^\star(\bar a, \bar b)$ at the equilibrium, for given price signals $\bar a$, $\bar b>0$ (dashed dotted coloured lines). The shaded blue area denotes the union over all the $25$ experiments performed.}
	\label{fig:normalized_demand}
\end{figure}

\subsection{Motivating example revisited}
Consider again the indirect control problem for smart grids presented in \S \ref{subsec:motivating_example}.
A \gls{DSO} or an energy retailer (i.e., an external observer) wants to make accurate forecasts on the aggregate electricity consumption of end-users in response to price-signals, aimed at enabling flexibility offered by the users themselves. We want to analyze the effect of price signals $a$, $b >0$ on the aggregate consumption $\sigma(\bs x^\star(a,b))$ of a fleet of \glspl{EV} over a certain period $T$, and the possibility to unlock the so-called ``valley filling'' phenomenon \cite{ma2011decentralized}.
We then consider the following collection of optimization problems:
\begin{equation}\label{eq:single_prob_EV}
	\forall i \in \mc N : \left\{
	\begin{aligned}
		& \underset{x_i}{\textrm{min}} &&  q_i x^\top_i x_i+c_i^\top x_i + (a (\sigma(\bs x) + d) + b \bsone_T)^\top x_i\\
		& \textrm{ s.t. } && \bsone_T^\top x_i \ge \gamma_i,~x_i \in [0, \bar x_i]^T,\\
		&&& \sigma(\bs x) \le \bar c,\\
	\end{aligned}
	\right.
\end{equation}
where we have used the values reported in Tab.~\ref{tab:sim_val} to run numerical experiments. In particular, we have run $25$ instances of \eqref{eq:single_prob_EV} obtained from all the possible combinations of five values for both $a$ and $b$ chosen equally spaced within the set $[0.02,0.8]$. In all the considered cases, we verified ex-post that the minimum eigenvalue of the Hessian matrix $H$ associated with the cost in \eqref{eq:constrained-LS} satisfies $\lambda_\textrm{min}(H) \ge 4.8\times10^{-2}$.

Then, Fig.~\ref{fig:ev_convergence} reports the averaged convergent behaviour of the sequence of queries $\{\hat{\bs x}^k\}_{k\in\N}$ produced by Algorithm~\ref{alg:learning_embedded_BR}. Note that we usually obtain an equilibrium $\bs x^\star$, computed through a standard extragradient type method \cite{solodov1996modified}, in less than the $K=200$ iterations allocated. In Fig.~\ref{fig:normalized_demand}, instead, we show the effect that different price signals have on the aggregate charging strategy of the fleet of \glspl{EV}. Specifically, we note that acting on the parameters $a$, $b$  produces a fine control of the aggregate \glspl{EV} consumption, increasing demand during off-peak periods, thereby producing a desirable valley-filling phenomenon. 

Therefore, our active learning procedure represents a key tool for a \gls{DSO} or an energy retailer in forecasting the aggregate electricity consumption of private customers in response to price-signals, in order to enable for the flexibility offered by the users themselves, or eventually maximize their profits. 

\subsection{Generalized Nash games}
For this multi-agent decision-making problem we consider examples from \cite{salehisadaghiani2017admm,facchinei2009penalty} to cover several cases of interest.

Specifically, \cite[Ex.~1]{salehisadaghiani2017admm} considers $N=10$ agents with cost functions $J_i(x_i, \bs x_{-i}) = N(1+(i-1)/2) x_i - x_i(60 N - \bsone^\top_N \bs x)$, while $\Omega = [7, 100]^N$. 
\begin{figure}[!t]
	\centering
	\ifTwoColumn
\else
	\includegraphics[width=.6\columnwidth]{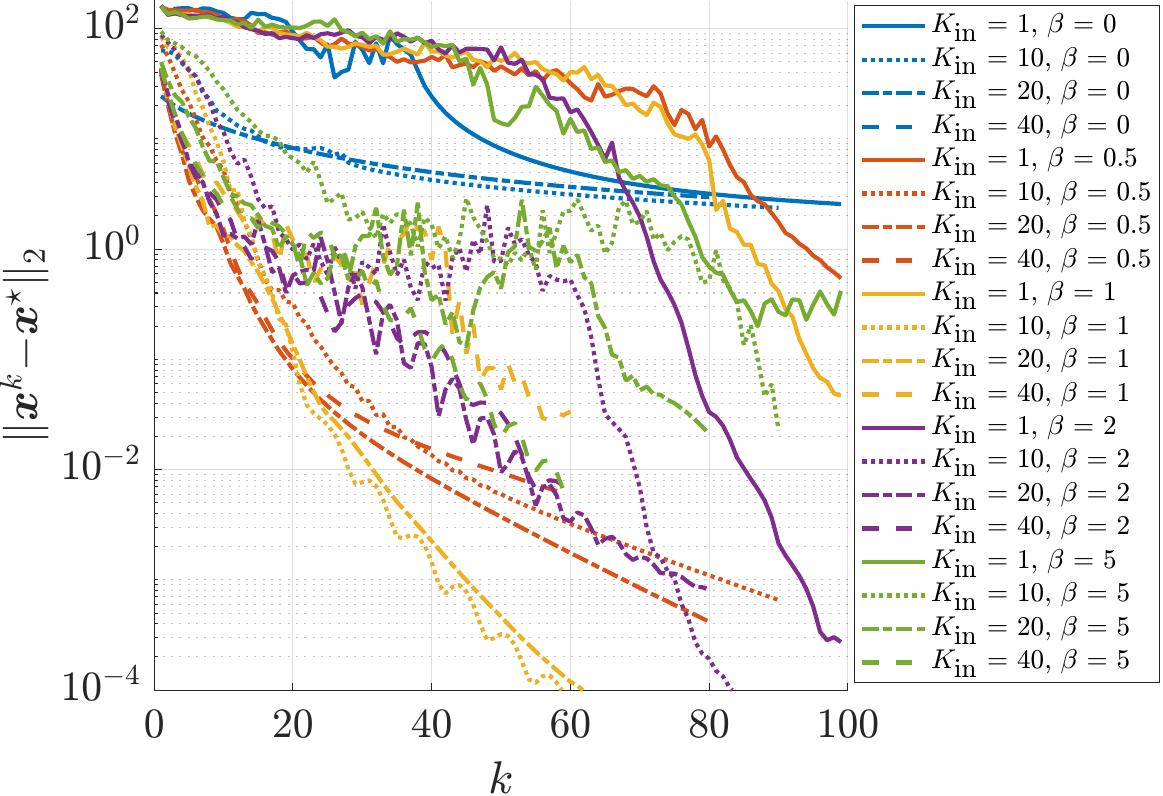}
	\fi
	\caption{Hyperparameters analysis on the Nash equilibrium problem formalized in \cite[Ex.~1]{salehisadaghiani2017admm}.}
	\label{fig:hyper_salehisadaghiani}
\end{figure}
We first perform a numerical analysis to set the two main hyperparameters affecting the execution of Algorithm~\ref{alg:learning_embedded_BR}, namely the learning coefficient $\beta$ and the number $K_{\textrm{in}}$ of iterations during the data collection phase. The selected values will be adopted also for the remaining numerical instances considered in this subsection. Fig.~\ref{fig:hyper_salehisadaghiani} illustrates the impact those hyperparameters have on the convergence of Algorithm~\ref{alg:learning_embedded_BR} to an equilibrium of the underlying game (computed through a standard extragradient type method \cite{solodov1996modified}), averaged over $20$ random initialization procedures for each coefficients combination. In particular, we have considered $\beta \in \{0, 0.5, 1, 2, 5\}$ and $K_{\textrm{in}} \in \{0.01, 0.1, 0.2, 0.4\} \cdot K$, with $K=100$. Different rules for
selecting $K_{\textrm{in}}$ can be adopted, e.g., choosing
$K_{\textrm{in}}$ as a function of the dimension $n_{-i}\!-\!1$ of $\theta_i$.

What we can infer from Fig.~\ref{fig:hyper_salehisadaghiani} is that while, on the one hand, it seems that some long enough random initialization procedure is indeed required (solid lines corresponding to $K=1$
are associated with the slowest behaviours), on the other hand, an exceedingly large learning rate (green and violet lines) does not appear to bring significant benefit in terms of rate of convergence. We will thus use $\beta=1$ and $K_{\textrm{in}}=\lceil K/10\rceil$.

\begin{figure}[!t]
	\centering
	\ifTwoColumn
\else
	\includegraphics[width=.6\columnwidth]{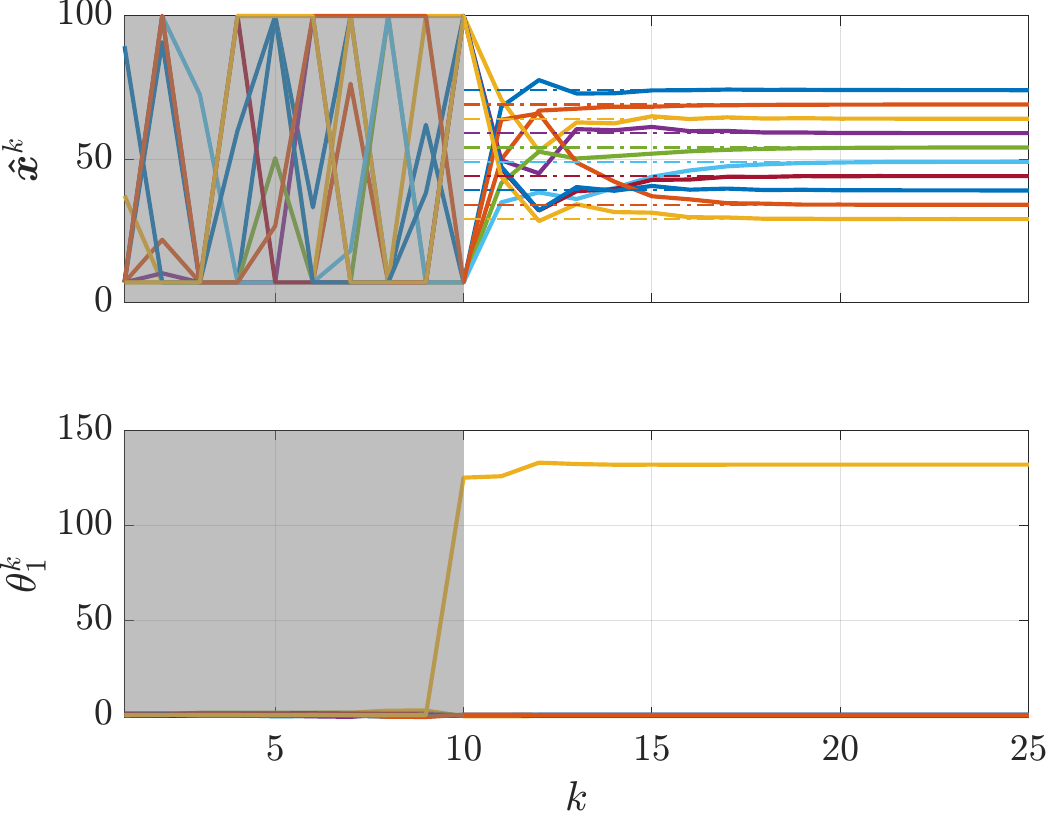}
	\fi
	\caption{Nash equilibrium problem formalized in \cite[Ex.~1]{salehisadaghiani2017admm}. The shaded regions correspond to the random data collection phase.
		Top: 
		sequence $\{\hat{\bs x}^k\}_{k\in\N}$ (solid lines) converging to an equilibrium (dashed lines). Bottom: convergent behaviour of $\{\theta_i^k\}_{k\in\N}$ to reconstruct the \gls{BR} mapping of agent $1$.
		Only the first 25 samples out of the allocated $K=100$ are reported.}
	\label{fig:salehisadaghiani}
\end{figure}
The performance of the active learning-based approach in predicting an equilibrium solution of the considered quadratic game with no coupling constraints is illustrated in Fig.~\ref{fig:salehisadaghiani}. In particular, the top plot shows the behaviours of the query points $\hat{\bs x}^k$ computed iteratively by the external entity to collect information on the agents' \gls{BR} mappings.
Remarkably, the approximation accuracy of the \gls{BR} mapping surrogates increases over the iterations, yielding $\lim_{k \to \infty} \|\hat{\bs x}^k - \bs x^k\|_2 = 0$, and therefore allowing the external entity to practically succeed in its prediction task in less than $20$ iterations. These considerations are motivated by the bottom plot in Fig.~\ref{fig:salehisadaghiani} that reports the convergent behaviour of $\theta_1^k$ (the other parameter vectors have a similar evolution), thus supporting Theorem~\ref{th:convergence} numerically.

\begin{figure}[!t]
	\centering
	\ifTwoColumn
\else
	\includegraphics[width=.6\columnwidth]{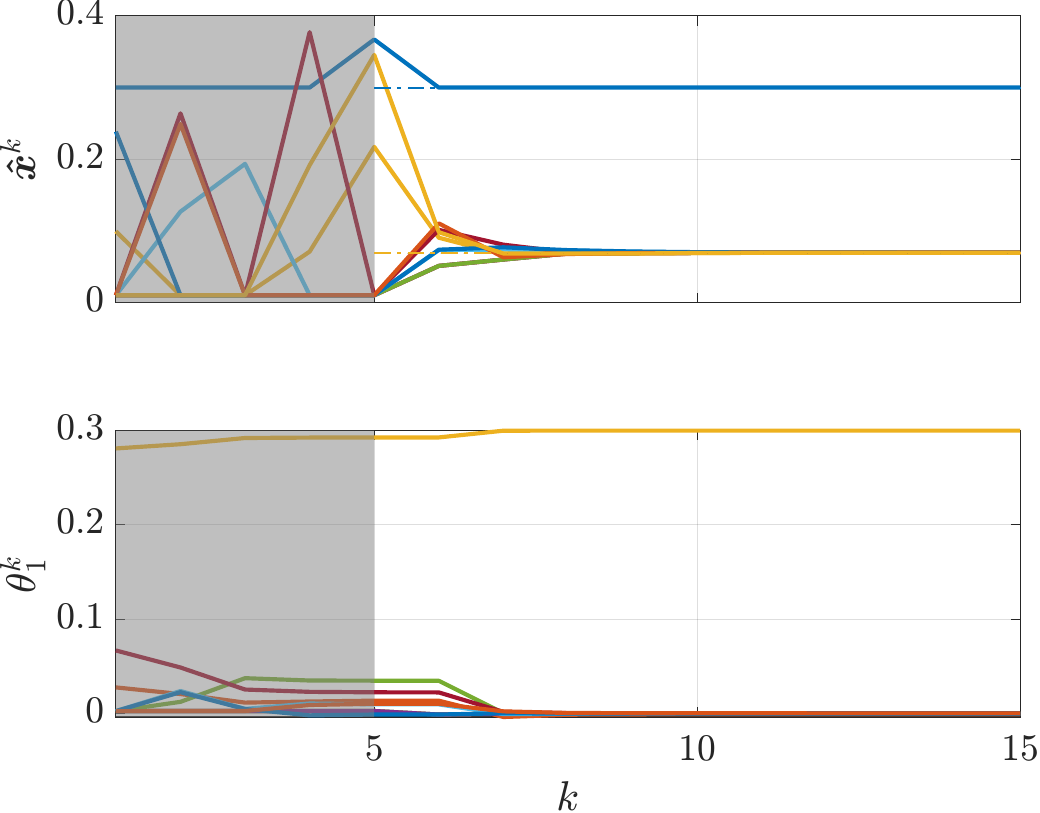}
	\fi
	\caption{\gls{GNEP} formalized in \cite[Ex.~1]{facchinei2009penalty}. The shaded regions correspond to the random data collection phase.
		Top: 
		sequence $\{\hat{\bs x}^k\}_{k\in\N}$ (solid lines) converging to an equilibrium (dashed lines). Bottom: convergent behaviour of the parameter vector reconstructing the \gls{BR} mapping of agent $1$.}
	\label{fig:facchinei_1}
\end{figure}

\begin{figure}[!t]
	\centering
	\ifTwoColumn
\else
	\includegraphics[width=.6\columnwidth]{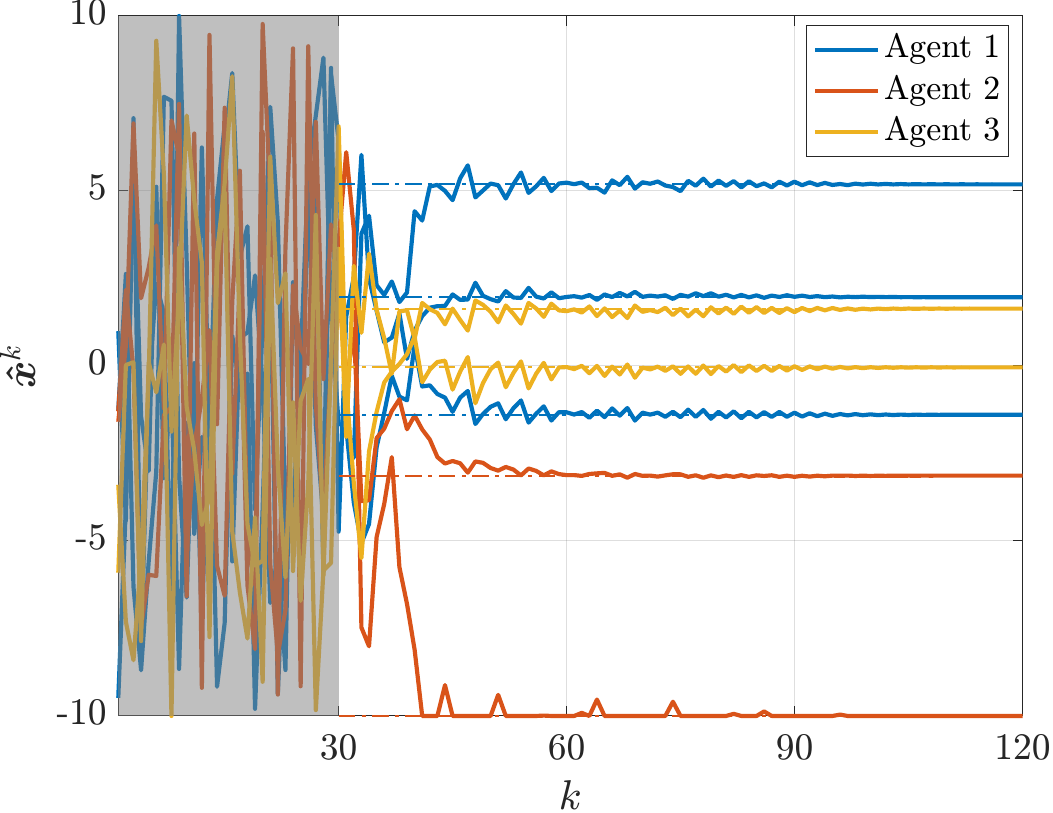}
	\fi
	\caption{\gls{GNEP} described in \cite[Ex.~3]{facchinei2009penalty}. The shaded region corresponds to the data collection phase ($K_{\textrm{in}}=30$). Sequence $\{\hat{\bs x}^k\}_{k\in\N}$ (solid lines) converging to a \gls{GNE} (dashed lines).}
	\label{fig:facchinei_3}
\end{figure}
Similar conclusions can be also drawn from Fig.~\ref{fig:facchinei_1} that illustrates the case in which Algorithm~\ref{alg:learning_embedded_BR} is applied to the example described in \cite[Ex.~1]{facchinei2009penalty} (in this case, $K=50$). Modelling the internet switching behaviour induced by selfish users, in this example each cost function reads as:
$$
	J_i(x_i, \bs x_{-i}) = -\frac{x_i}{\bsone^\top_N \bs x} \left(1 - \bsone^\top_N \bs x\right).
$$ 
Then, while the first agent's strategy is constrained so that $x_1 \in [0.3, 0.5]$, for all $j\neq1$ we have $x_j \in [0.01, 100]$. In addition, a standard linear constraint $\bsone^\top_N \bs x \le 1$ couples the strategies of the whole population of decision-makers, thus actually resulting into a \gls{GNEP}. An example with multidimensional decision vectors is instead the one illustrated in Fig.~\ref{fig:facchinei_3}, which amounts to the \gls{GNEP} described in \cite[Ex.~3]{facchinei2009penalty} and involving $N=3$ agents with three, two and two variables, respectively. Even though the point $\{\hat{\bs x}^k\}_{k \in \N}$ converges to is different from the \gls{GNE} reported in \cite[Ex.~3]{facchinei2009penalty}, it still coincides with an equilibrium of the underlying \gls{GNEP}, being a fixed point of the  \gls{BR} mappings.

\begin{figure}[!t]
	\centering
	\ifTwoColumn
\else
	\includegraphics[width=.6\columnwidth]{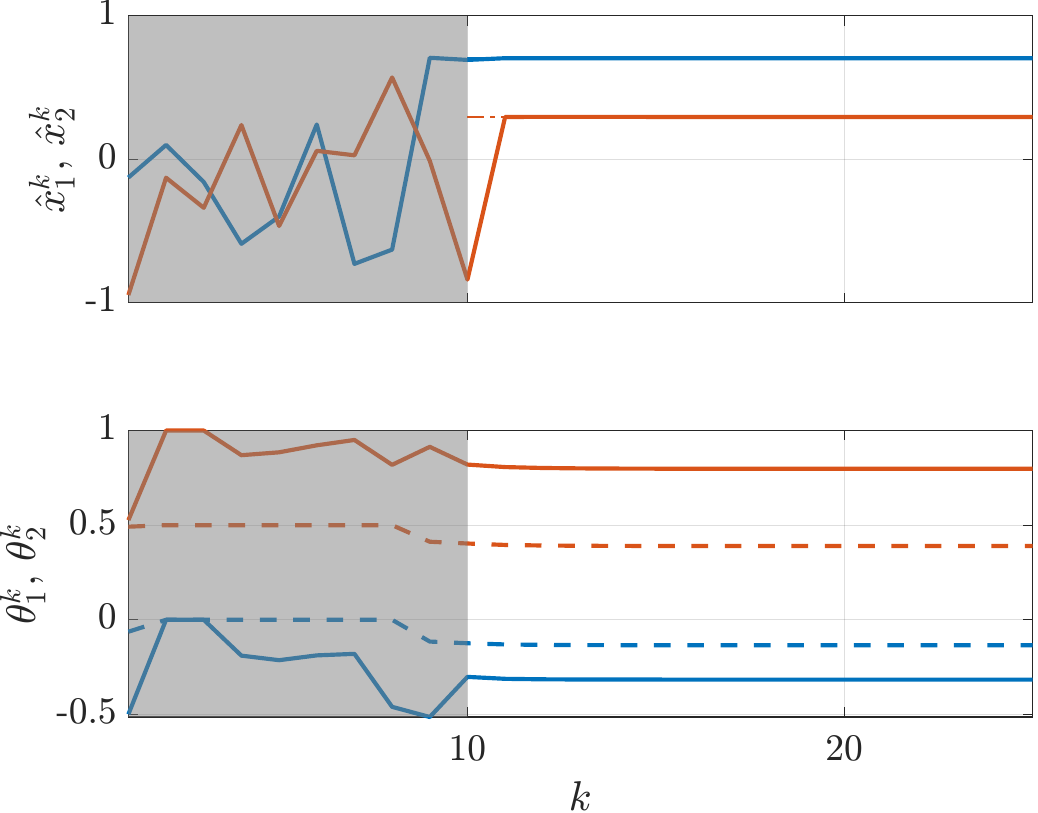}
	\fi
	\caption{\gls{GNEP} formalized in \cite[Ex.~11]{facchinei2009penalty}. The shaded region corresponds to the random data collection phase ($K=100$).
	Top: 
	Sequence $\{\hat{\bs x}^k\}_{k\in\N}$ (solid lines) converging to the normalized equilibrium (dashed lines). Bottom: convergent behaviour of the parameter vectors reconstructing the \gls{BR} mapping for both agents (same colour, solid lines correspond to the first element of each $\theta_i$, while dashed lines to the second).}
	\label{fig:facchinei_11}
\end{figure}
\begin{figure}[!t]
\centering
\ifTwoColumn
\else
\includegraphics[width=.6\columnwidth]{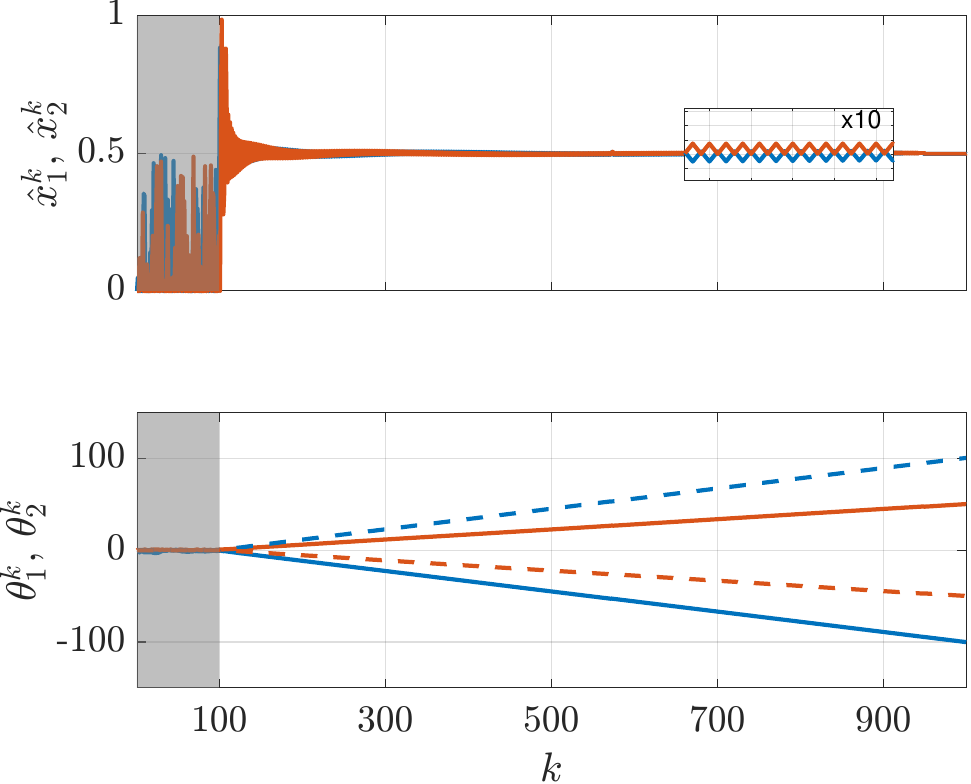}
\fi
\caption{Nash equilibrium problem with no equilibria. The shaded region corresponds to the random data collection phase ($K_{\textrm{in}}=100$, $K=1000$). Top: 
	Sequence $\{\hat{\bs x}^k\}_{k\in\N}$ obtained by the external entity. Bottom: diverging behaviour of the parameter vectors reconstructing the \gls{BR} mapping for both agents (same colour, solid lines correspond to the first element of each $\theta_i$, while dashed lines to the second one).}
\label{fig:-2}
\end{figure}
We conclude this part by discussing two simple, yet significant, examples involving two agents only. The first one consists in the \gls{GNEP} in \cite[Ex.~11]{facchinei2009penalty} featuring an infinite number of equilibria $\set{\col(\alpha, 1-\alpha)}{\alpha \in[1 / 2,1]}$, for which Fig.~\ref{fig:facchinei_11} shows that Algorithm~\ref{alg:learning_embedded_BR} is actually able to return one of those solutions -- specifically that coinciding with $\alpha=0.7$. 
After executing such an example $1000$ times, we have noticed that Algorithm~\ref{alg:learning_embedded_BR} always returns the same equilibrium corresponding to $\alpha=0.7$ in all those instances in which it has converged.
On the contrary, the second example considered does not have any Nash equilibrium due to the lack of quasiconvexity in the agents' cost functions, which are taken as $J_1(x_1,x_2) = -x_1^2 + 2 x_1 x_2$ and $J_2 = x_2 - 2 x_1 x_2$, $x_i \in [0,1]$ for $i \in \{1,2\}$. Figure~\ref{fig:-2} reports the numerical results for this example, where it is evident the non-convergent behaviour of the main quantities involved, particularly the parameter vectors $\theta_i$. According to the discussion in \S \ref{subsec:discussion}, in this case the sufficient conditions established in Theorem~\ref{th:convergence} 
do not hold, and therefore no certificate for the existence of Nash equilibria is provided, which is indeed the case.
Finally, Table~\ref{tab:min_eig} shows that, in all those numerical instances considered for which convergence of the parametric estimates happens, also the sufficient condition for the uniqueness of the solution in $\mc M(\tilde\theta)$ is met (namely, positive definiteness of the Hessian matrix $H$ of
the cost function, along with convexity of $\Omega$), thus verifying the requirements in Theorem~\ref{th:convergence} numerically.

\begin{table}[t!]
	\caption{Minimum eigenvalue of the Hessian matrix of \eqref{eq:minimizers}}
	\label{tab:min_eig}
	\begin{center}
		\begin{tabular}{llllll}
			\toprule
			  &  Fig.~\ref{fig:salehisadaghiani} & Fig.~\ref{fig:facchinei_1} & Fig.~\ref{fig:facchinei_3} &  Fig.~\ref{fig:facchinei_11} & Fig.~\ref{fig:-2}\\
			\midrule
			$\lambda_\textrm{min}(H)$ &$0.11$ & $0.16$ & $0.001$ & $0.65$ & $\ast$ \\
			\bottomrule
		\end{tabular}
	\end{center}
\end{table}

\subsection{Multi-agent feedback controller synthesis}\label{subsec:dec_con_sim}
The control application described in \S \ref{sec:applications} is slightly more challenging, mostly because the conditions for the stack of the (highly nonlinear) action-reaction mappings \eqref{eq:action_reaction_LQR} to admit a stationary action profile are not clear. 
For this reason, we selected $10$ random numerical instances on which Algorithm~\ref{alg:learning_embedded_BR} returned a fixed point of each $f_i(\cdot)$ in \eqref{eq:action_reaction_LQR} after just one execution, thus ensuring the existence of a stationary action profile according to Corollary~\ref{cor:existence} (Fig.~\ref{fig:ex_1} reports just one converging instance of the first example for agent $1$).
In particular, we generated random, discrete-time \gls{LTI} test models with unstable eigenvalues, satisfying the observability and reachability conditions discussed in \S \ref{subsec:dec_con}. Given $N=3$ agents with $n_{u_i}=1$ for simplicity, we have $n_z\sim\mc U(N,3N)$, $n_y \sim \mc U(N,2N)$ and $n_{y_i} \sim \mc U(2,n_y)$, while weights $r_i \sim \mc U(1,10)$, $Q_i = W W^\top$ and $W \sim \mc U(0,1)$, $i \in \mc N$. 

\begin{figure}[!t]
	\centering
	\ifTwoColumn
\else
	\includegraphics[width=.6\columnwidth]{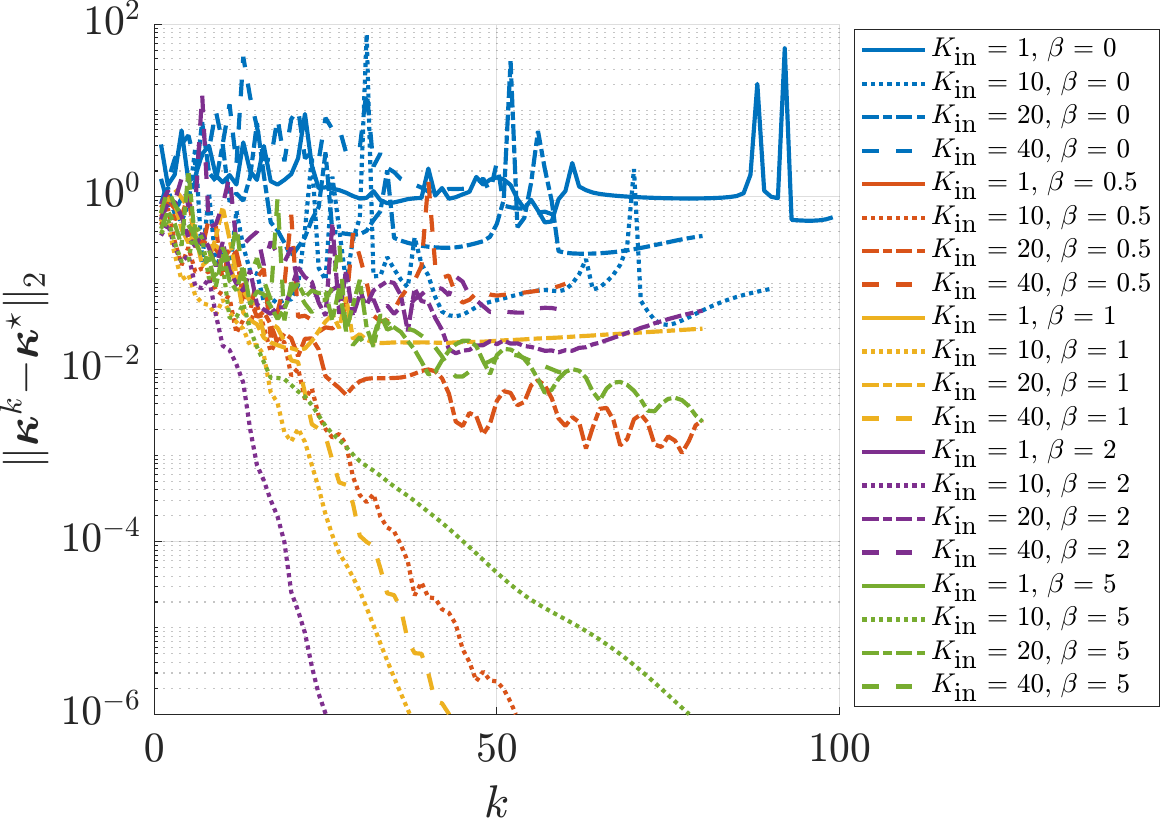}
	\fi
	\caption{Hyperparameters analysis on the first randomly drawn multi-agent feedback controller synthesis problem.}
	\label{fig:hyper_dec_controrl}
\end{figure}

\begin{figure}[!t]
	\centering
	\ifTwoColumn
\else
	\includegraphics[width=.6\columnwidth]{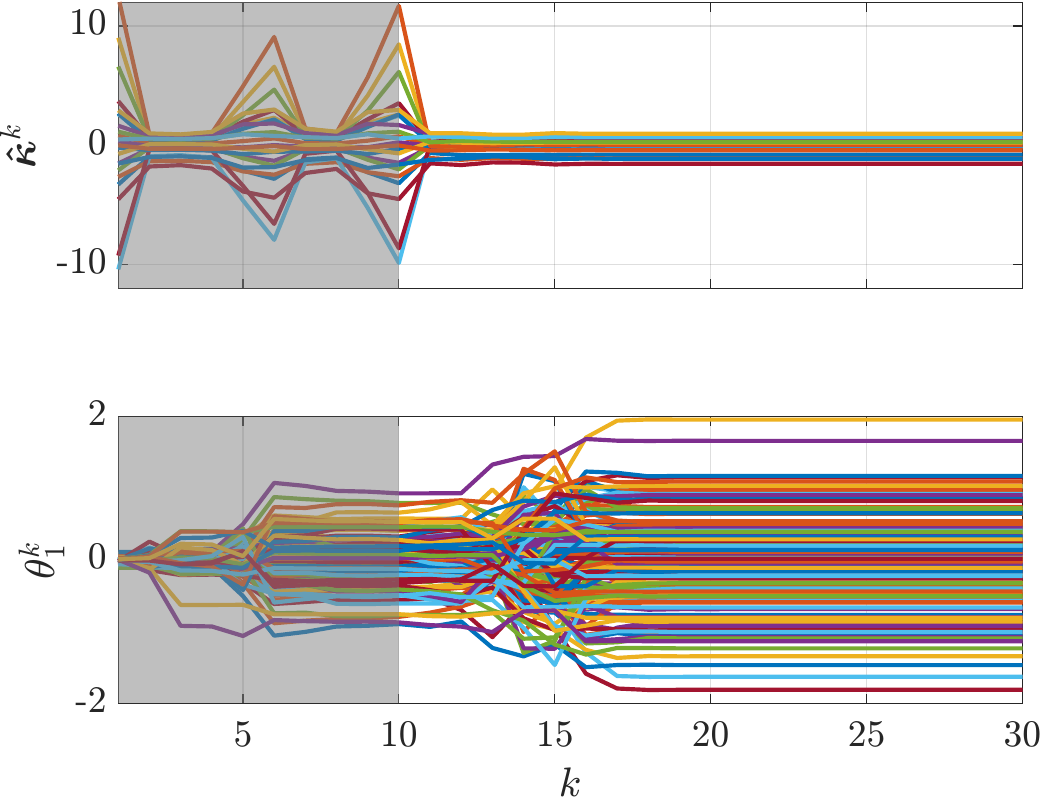}
	\fi
	\caption{First randomly drawn multi-agent feedback controller synthesis problem. The shaded region corresponds to the random data collection phase ($K=100$). Top: 
	Sequence $\{\hat{\bs x}^k\}_{k\in\N}$ converging to a stationary action profile. Bottom: convergent behaviour of the parameter vector reconstructing the action-reaction mapping \eqref{eq:action_reaction_LQR} of agent $1$.}
	\label{fig:ex_1}
\end{figure}

We started by analyzing the impact the hyperparameters have on the convergence of Algorithm~\ref{alg:learning_embedded_BR} when applied to the first randomnly generated instance. We replicate precisely the same approach as described in the previous subsection, i.e., $K=100$, average over $20$ executions of Algorithm~\ref{alg:learning_embedded_BR} for coefficients combination $\beta \in \{0, 0.5, 1, 2, 5\}$ and $K_{\textrm{in}} \in \{0.01, 0.1, 0.2, 0.4\} \cdot K$. The results in Fig.~\ref{fig:hyper_dec_controrl} show that, unlike the \gls{GNEP} setting, for this application the convergence rate is more sensible to the learning rate $\beta$ rather than the length of the random initialization phase (the blue lines do not even converge indeed). We thus set $\beta=2$ and $K_\textrm{in} = 0.1 K$ to conduct the numerical simulations of this subsection.

The random initialization phase is here performed by presenting to the agents centralized \gls{LQR} gains obtained with output and input weights $Q=I_{n_y}$ and $R=I_{N}$, and after perturbing the original dynamical matrix. Moreover, to avoid the emptiness of \eqref{eq:action_reaction_LQR} due to the unboundedness of $\kappa_i$, for some $\bs \kappa_{-i}$ the external entity computes points $\hat{ \bs \kappa} \in \hat\Omega= \Omega = [-20, 20]^{N n_z}$.

As already spotted out in the comments following Fig.~\ref{fig:hyper_salehisadaghiani}, from our numerical experience we have found that, given a numerical instance enjoying the existence of a stationary action profile, Algorithm~\ref{alg:learning_embedded_BR} may not converge at each execution. With the application considered in this subsection, however, this behaviour is even more pronounced, possibly also in view of the highly nonlinear structure of the underlying mappings. These considerations hence motivate us in performing a statistical analysis to shed further light on this fact, and the last column of Table~\ref{tab:num_res} reports the percentage of convergent runs over $1000$ executions of Algorithm~\ref{alg:learning_embedded_BR}. In all the numerical simulations in which convergence of the parametric estimates happens -- possibly to different points -- we observe that: i) Algorithm~\ref{alg:learning_embedded_BR} returns a fixed point of the stack of the action-reaction mappings, thus verifying numerically the theory in \S \ref{sec:learning_proc}; and, quite interestingly, ii) the point it returns is always the same. Finally, to populate the last column of Table~\ref{tab:num_res}, for each numerical example we have taken the smallest eigenvalue of the Hessian matrix for \eqref{eq:constrained-LS} among the minimum ones $\lambda^i_\textrm{min}(H)$ in all those cases in which the convergence of the parametric estimates happened. Such a column indeed shows that the condition on $\mc M(\tilde\theta)$ of being a singleton is always met.


\begin{table}[t!]
	\caption{Multi-agent feedback controller synthesis -- Numerical results}
	\label{tab:num_res}
	\begin{center}
		\begin{tabular}{llllll}
			\toprule
			Example  &  $n_z$ & $n_y$ & $n_{y_i}$ &  $\%_{\textrm{conv}}$ & $\underset{i}{\textrm{min}}~\{\lambda^i_\textrm{min}(H)\}$\\
			\midrule
			$1$ & $8$ & $6$ & $\{5,2,5\}$ & $60$ & $5.39\times10^{-4}$\\
			\midrule
			$2$ & $7$ & $4$ & $\{3,2,4\}$ & $49.3$ & $1.23\times10^{-3}$\\
			\midrule
			$3$ & $4$ & $6$ & $\{2,4,2\}$ & $87.6$ & $4.29\times10^{-4}$\\
			\midrule
			$4$ & $5$ & $6$ & $\{4,5,5\}$ & $82.5$ & $1.76\times10^{-4}$\\
			\midrule
			$5$ & $6$ & $5$ & $\{2,5,3\}$ & $98.1$ & $1.05\times10^{-4}$\\
			\midrule
			$6$ & $4$ & $6$ & $\{2,3,3\}$ & $89.5$ & $2.59\times10^{-4}$\\
			\midrule
			$7$ & $8$ & $6$ & $\{2,5,4\}$ & $94.9$ & $6.80\times10^{-3}$\\
			\midrule
			$8$ & $4$ & $3$ & $\{3,3,2\}$ & $98.6$ & $1.85\times10^{-4}$\\
			\midrule
			$9$ & $5$ & $6$ & $\{2,5,2\}$ & $62.9$ & $1.79\times10^{-4}$\\
			\midrule
			$10$ & $8$ & $5$ & $\{4,5,2\}$ & $94.7$ & $3.54\times10^{-3}$\\
			\bottomrule
		\end{tabular}
	\end{center}
\end{table}

\begin{figure*}
	\centering
	\begin{subfigure}[b]{0.5\columnwidth}
		\centering
		\includegraphics[width=\columnwidth]{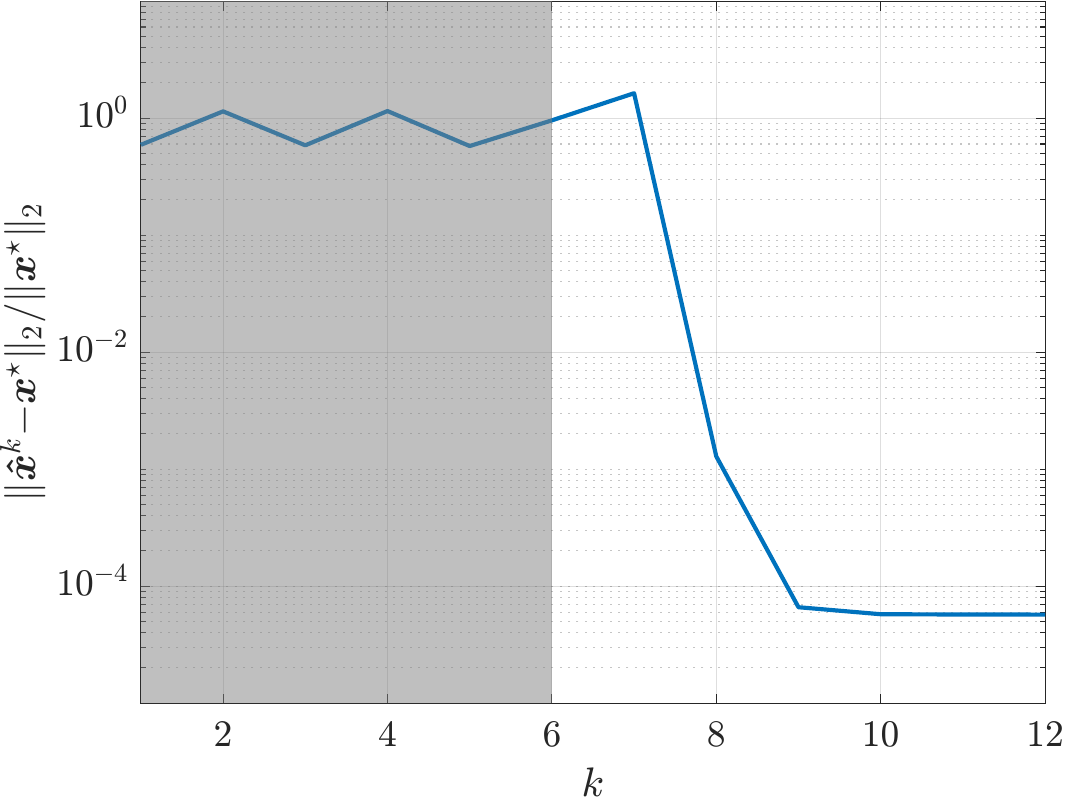}
		\caption{}
		\label{fig:a}
	\end{subfigure}
	\hfill
	\begin{subfigure}[b]{0.5\columnwidth}
		\centering
		\includegraphics[width=\columnwidth]{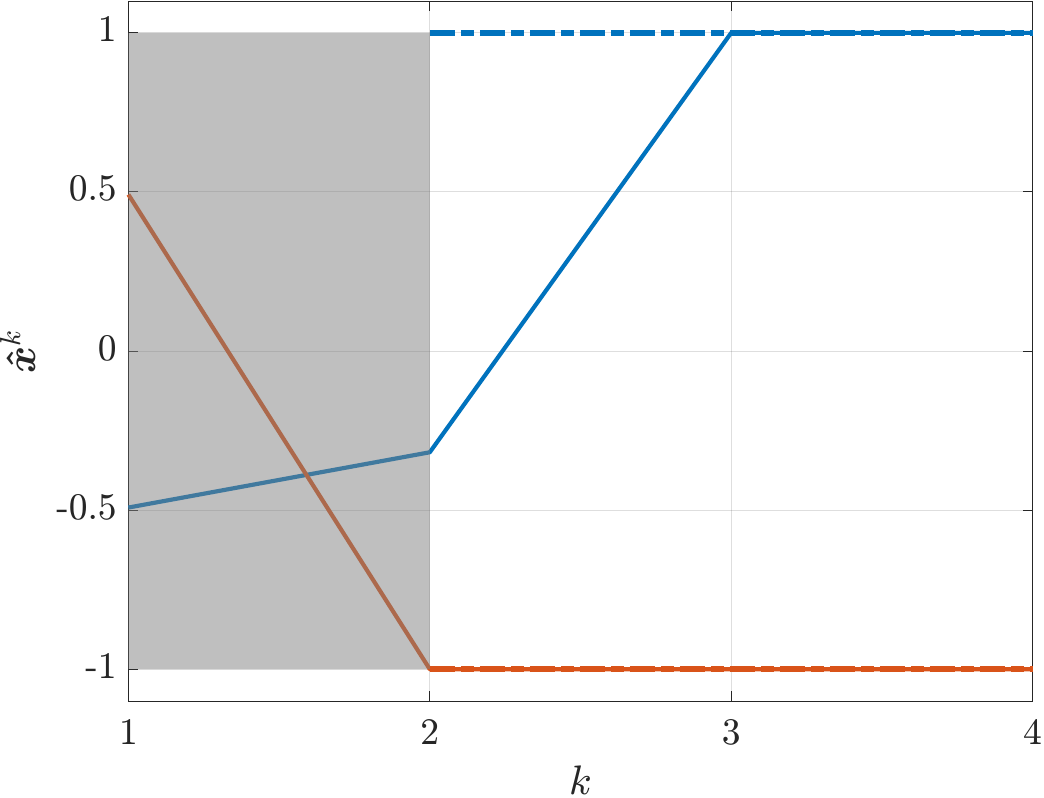}
		\caption{}
		\label{fig:b}
	\end{subfigure}
	\hfill
	\begin{subfigure}[b]{0.5\columnwidth}
		\centering
		\includegraphics[width=\columnwidth]{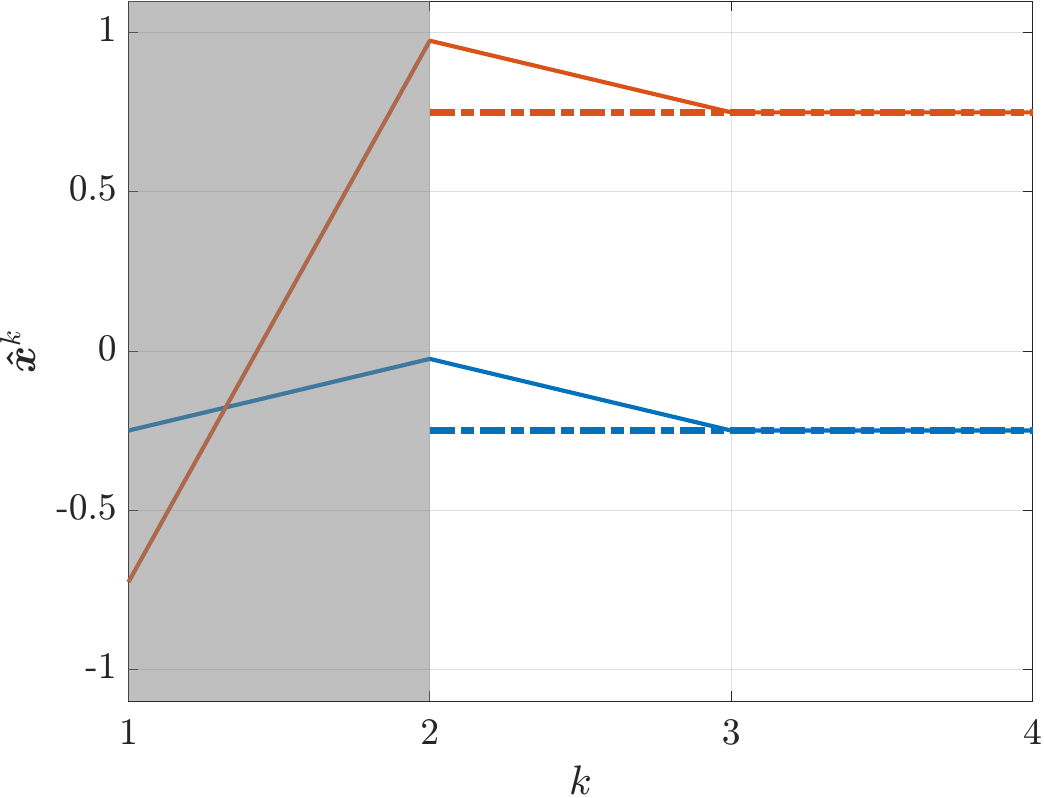}
		\caption{}
		\label{fig:c}
	\end{subfigure}
	\hfill
	\begin{subfigure}[b]{0.5\columnwidth}
		\centering
		\includegraphics[width=\columnwidth]{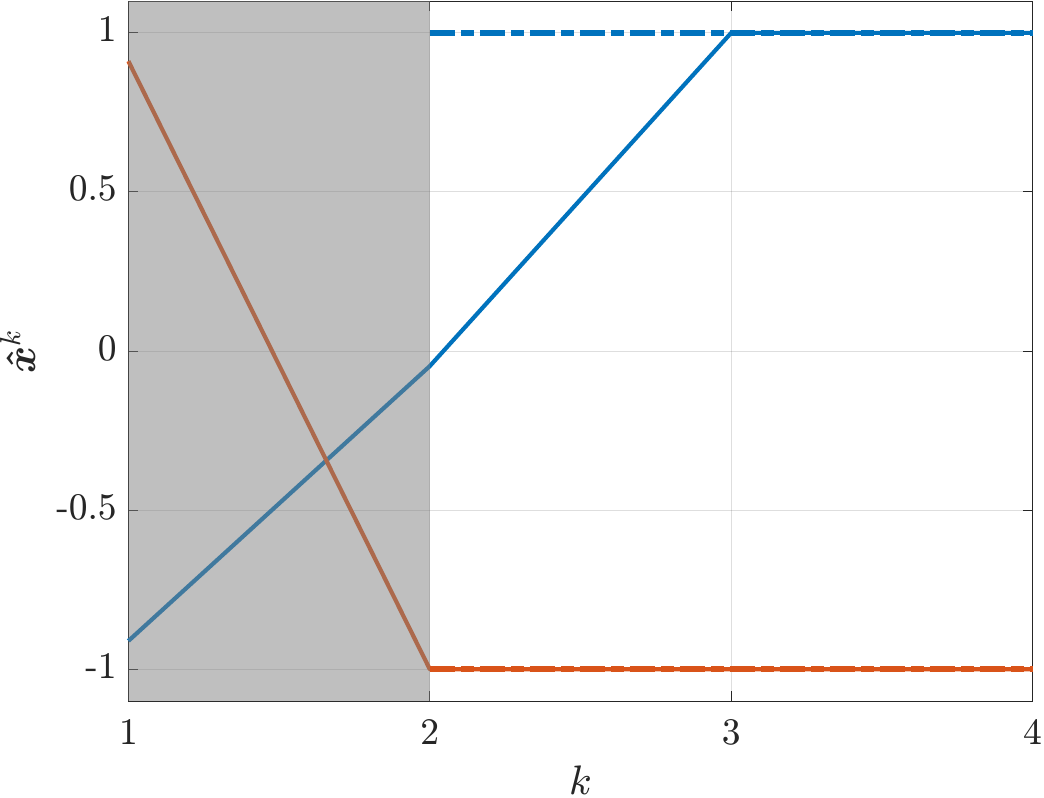}
		\caption{}
		\label{fig:d}
	\end{subfigure}
	\caption{Convergence behaviour exhibited by Algorithm~\ref{alg:learning_embedded_BR} in four of the examples provided with \texttt{TensCalc} toolbox: (a) Two-player zero-sum game with $n_1=n_2=200$ and quadratic cost functions $J_1(x_1,x_2)=\|x_1\|^2+2(x_2-2\cdot\bsone_{200})^\top x_1$, $J_2(x_2,x_1)=9 \|x_2\|^2-2(x_1-2 \cdot \bsone_{200})^\top x_2$; (b) Min-max problem with scalar decision variables, $J_1(x_1,x_2)=(x_1+x_2)^2$, $J_2(x_2,x_1)=-(x_2+x_1)^2+2(x_2+2)^2$, and $|x_2|\le 1$; (c) Min-max problem with scalar decision variables, $J_1(x_1,x_2)=(x_1+x_2+1)^2$, $J_2(x_2,x_1)=-(x_2+x_1+1)^2+2x_2^2$, and $|x_1|\le 0.25$; (d) Min-max problem with scalar decision variables, $J_1(x_1,x_2)=(x_1+x_2)^2$, $J_2(x_2,x_1)=-(x_2+x_1)^2+2(x_2+2)^2$, $|x_1|\le 2$, and $|x_2|\le1$.}
	\label{fig:comparison}
\end{figure*}

\subsection{Comparison results}
We now compare the performance of Algorithm~\ref{alg:learning_embedded_BR} against those offered by \texttt{TensCalc} \cite{hespanha2022tenscalc}, a centralized state-of-the-art method for \gls{GNE} computation. We remark that, unlike our approach, \texttt{TensCalc} requires a fully centralized scenario where the agents' cost functions $J_i$ and local constraint sets $\mathcal{X}_i$ have to be known by the central entity.
We thus consider four examples provided with the toolbox associated to \texttt{TensCalc}\footnote{Available at {\tt \href{https://github.com/hespanha/tenscalc}{https://github.com/hespanha/tenscalc}}.}, namely a zero-sum game and three different min-max problems, illustrated and described in Fig.~\ref{fig:comparison}.

Specifically, Fig.~\ref{fig:comparison} reports just one numerical instance on top the $50$ generated to compare the average computational time of the two approaches. There, one can appreciate how our active learning scheme produces a convergent sequence
to the \gls{NE} (computed with \texttt{TensCalc}).
In the first example in Fig.~\ref{fig:comparison}.(a), the random initialization phase and the four iterations required to converge take $2.2$~[s], as opposed to \texttt{TensCalc} that provides a solution in $1.3$~[ms]. For the examples in Fig.~\ref{fig:comparison}.(b)--(d), instead, Algorithm~\ref{alg:learning_embedded_BR} shows a fast convergence, ranging from $34.2$~[ms] to $42.1$~[ms], while \texttt{TensCalc} is generally slower ($0.3$ to $0.34$~[s]).
In all cases, note that \texttt{TensCalc} needs to ``generate'' a solver, a procedure lasting for around $4$ seconds on the laptop employed.

\section{Conclusion and outlook}
We have proposed an active learning scheme allowing an external entity to actively learn faithful local
surrogates of action-reaction mappings, privately held by a population of interacting agents, in order to find a stationary action profile of the underlying multi-agent interaction process. We showed in numerical
experiments the practical effectiveness and versatility of the procedure on a number of typical competitive multi-agent decision-making and control problems.

The proposed approach paves the way to numerous extensions. Looking at the specific problem of learning a \gls{GNE} in \glspl{GNEP}, for instance, given the tight connection between an equilibrium solution and a minimizer of the Nikaido-Isoda function \cite{nikaido1955note}, one may design a tailored procedure along the line of Algorithm~\ref{alg:learning_embedded_BR} to locally approximate such a function. 
Finally, we remark that problem~\eqref{eq:minimizers} is based completely on
\emph{exploiting} the current predictors $\hat f_i$. In alternative to the initial random exploration phase, one could also investigate possible solutions for better  \emph{exploring} the decision set $\Omega$, 
as done for example in reinforcement learning~\cite{SB18} and surrogate-based global
optimization~\cite{Bem20}, and analyze its potential benefits.


\appendix
\section{Appendix}\label{sec:proofs}
\subsection{Technical proofs}

\emph{Proof of Lemma~\ref{lemma:convexity_fk}:}
Let $\bar{r}:\R^n\times\R^p\to\bar\R$ be defined by $\bar{r}(x,\theta)\eqdef
r(x,\theta)+\iota_\mc X(x)$.
Since $\mc X$ is a convex set and $r$ is a convex function, then $\bar{r}$ is also a convex function for each given $\theta\in\R^p$. By applying \cite[Th.~2.6]{rockafellar2009variational},
it follows that $\mc M(\theta)=\textrm{argmin}_{y\in \R^n}~\bar{r}(y,\theta)$
is a convex set for all $\theta\in\R^p$.

To show continuity of $r^\star$ at a generic $\tilde\theta\in\R^p$, we need to prove
that
$
\lim_{k\rightarrow\infty}r^\star(\theta^k)=r^\star(\tilde\theta),
$
for any sequence $\{\theta^k\}_{k\in\N}$ such that $\lim_{k\rightarrow\infty}\theta^k=\tilde\theta$.
Consider such a generic sequence and let $x^k\in \mc M(\theta^k)$ and $\tilde x\in \mc M(\tilde\theta)$. 
In view of Assumption~\ref{ass:convexity-2}, $\theta\mapsto r(x^k,\theta)$ is also convex and differentiable \gls{wrt} $\theta$ and $\tilde x\in \mc M(\tilde\theta)$, i.e., $\tilde x$ is a global minimizer of $r(\cdot,\tilde\theta)$, and therefore we have that
\[
\begin{split}
	r^\star(\theta^k)=r(x^k,\theta^k)&\geq r(x^k,\tilde\theta)+\frac{\partial r(x^k,\theta)}{\partial \theta}\Big\rvert_{\theta=\tilde\theta}^\top(\theta^k-\tilde\theta)\\
	&
	\geq r(\tilde x,\tilde\theta)+\frac{\partial r(x^k,\tilde\theta)}{\partial \theta}\Big\rvert_{\theta=\tilde\theta}^\top(\theta^k-\tilde\theta)\\
	&\geq r(\tilde x,\tilde\theta)-D(\tilde\theta)\|\theta^k-\tilde\theta\|,
\end{split}
\]
where the last inequality holds since the partial derivative $\partial r(x,\theta)/\partial \theta$ evaluated in $\tilde\theta$ is bounded \gls{wrt} $x$, with bound $D(\tilde\theta)\eqdef\textrm{sup}_{x\in \mc X}\left|
\tfrac{\partial r(x,\theta)}{\partial \theta}\Big\rvert_{\theta=\tilde\theta}\right|\in\R$. Therefore, we obtain:
\begin{equation}
	r(\tilde x,\tilde\theta)-D(\tilde\theta)\|\theta^k-\tilde\theta\|\leq r^\star(\theta^k)=r(x^k,\theta^k)\leq 
	r(\tilde x,\theta^k)
	\label{eq:f-ineq}
\end{equation}
where the last inequality follows by the optimality of $x^k$.
Since $\theta\mapsto r(x,\theta)$ is continuous and $\lim_{k\rightarrow\infty}\theta^k=\tilde\theta$, we have that $\lim_{k\rightarrow\infty} r(\tilde x,\theta^k)=r(\tilde x,\tilde\theta)$, and the proof concludes by applying the squeeze theorem to \eqref{eq:f-ineq}.
\hfill\qedsymbol

\smallskip

\emph{Proof of Lemma~\ref{lemma:lsc_level-bounded}:}
Lower semicontinuity of $\bar{r}(x,\theta)$ readily follows by the continuity of $r(x,\theta)$ at any given point $(\tilde x,\tilde \theta)\in\R^n\times\R^p$, as
\[
\begin{aligned}
	\lim_{(x,\theta)\rightarrow(\tilde x,\tilde\theta)}\inf~\bar{r}(x,\theta) &=
	\lim_{\varepsilon\rightarrow 0}\left[
	\inf_{(y,\xi)\in \mc B_\varepsilon(\tilde x,\tilde\theta)} r(y,\xi)\right]\\
	&=
	\left\{
	\begin{array}{lll}
		r(\tilde x,\tilde\theta)&\mbox{if}&\tilde x\in \mc X\\
		+\infty& \mbox{if} & \tilde x\not\in \mc X
	\end{array}
	\right.
\end{aligned}
\]
which in turn implies that $\lim_{(x,\theta)\rightarrow(\tilde x,\tilde\theta)}\inf~\bar{r}(x,\theta)\geq \bar{r}(\tilde x,\tilde\theta)$ for all $(\tilde x,\tilde\theta)\in\R^n\times\R^p$. To show level-boundedness, note that $\mc X$ is bounded and nonempty, and therefore, for all $\alpha\in\R$, $\set{x\in\R^n}{\bar{r}(x,\theta)\leq\alpha}\subseteq \mc X$ for all $\theta\in\R^p$. Thus, by setting $\mc S=\mc X$ and $g = \bar{r}$ in Definition~\ref{def:level-bounded}, the claim follows.
\hfill\qedsymbol

\smallskip

\emph{Proof of Lemma~\ref{lemma:convergence}:}
By \cite[Th.~2.6]{rockafellar2009variational}, each vector $x^k$ in~\eqref{eq:xk} is uniquely defined since $\tfrac{1}{2}\|\cdot\|_2^2$ is strictly convex, proper because
$\mc X$ is nonempty and therefore, by Lemma~\ref{lemma:convexity_fk},
$\mc M(\theta^k)$ is a convex set. Moreover,
since $x\mapsto r(x,\theta)$ is continuous for all $\theta\in\R^p$ and $\mc X$ is compact,
by Lemma~\ref{lemma:lsc_level-bounded} the function $\bar{r}(x,\theta)$, defined by
$\bar{r}(x,\theta) = r(x,\theta)+\iota_\mc X(x)$, is lower semicontinuous \gls{wrt} $x$ and level-bounded in $x$ locally uniformly \gls{wrt} $\theta$.
Still in view of Lemma~\ref{lemma:convexity_fk}, we also have that $\lim_{k\rightarrow\infty}r^\star(\theta^k)=r^\star(\tilde\theta)$.
Thus, from \cite[Th.~1.17(b)]{rockafellar2009variational} it follows that the sequence $\{x^k\}_{k\in\N}$ generated by \eqref{eq:xk} is bounded and all its cluster points lie in $\mc M(\tilde\theta)$, which is however a singleton and contains $\tilde x$ only.

For the sake of contradiction, assume that $x^k\not\rightarrow\tilde x$ as $k\rightarrow\infty$.
Then, there exists some $\varepsilon>0$ and an infinite subsequence $\{x^k\}_{k\in \mc I}$ so that $\|x^k-\tilde x\|_2>\varepsilon$ for all $k\in \mc I$. Since $\{x^k\}_{k\in \mc I}$ is also bounded,
by the Bolzano-Weierstrass theorem there exists an infinite subsequence $\{x^k\}_{k\in \mc I'}$, with $\mc I'\subset \mc I$, 
such that $\lim_{k\rightarrow \infty, k\in \mc I'}x^k=\bar x$, with $\bar x\neq\tilde x$.
Since $\bar x$ is a cluster point of $\{x^k\}_{k\in\N}$, by relying on \cite[Th.~1.17(b)]{rockafellar2009variational} we have that $\bar x\in \mc M(\tilde\theta)$. This however is in conflict with the fact that $\mc M(\tilde\theta)$
is a singleton, i.e., $\mc M(\tilde\theta)=\{\tilde x\}$, and hence the
assumption $x^k\not\rightarrow\tilde x$ for $k\rightarrow\infty$ is contradicted, proving the statement.
\hfill\qedsymbol

\smallskip

\emph{Proof of Lemma~\ref{lemma:local_exactness}:}
Let us consider a single agent $i \in \mc N$, as the proof for the remaining ones is identical \emph{mutatis mutandis}. By making use of a contradiction argument, we will show that any of the parameters $\bar \theta_i \in \mc A_i \eqdef \mc A_i(\bar{x}_i, \bar{\bs x}_{-i})$ asymptotically belongs to the set of minimizers of $\xi_i \mapsto \frac{1}{k} \sum_{t = 1}^k \ell_i(x_i^t, \hat f_i(\hat{\bs x}^t_{-i}, \xi_i))$, and therefore it shall happen that $\bar{\theta}_i \in \mc A_i$. Recall, indeed, that the parameter update reads as $\theta_i^{k+1} = \textrm{argmin}_{\xi_i \in \R^{p_i}} \, \frac{1}{k} \sum_{t = 1}^k \ell_i(x_i^t, \hat f_i(\hat{\bs x}^t_{-i}, \xi_i))$, where equality follows by virtue of the global optimality assumed and the tie-break rule $\mc T(\cdot)$ in place. Thus, for the sake of contradiction, let us assume that $\lim_{k \to \infty} \theta_i^k = \tilde{\theta}_i$ with $\tilde{\theta}_i \notin \mc A_i$.

\sloppy To start we note that, by combining the definition of limit \cite[Def.~5.4]{thomson2001elementary} and the properties of the loss function stated in Standing Assumption~\ref{standing:learning_procedure}, along with those of the affine surrogates $\hat f_i$ described in \S \ref{subsec:convergence}, $\lim_{k \to \infty} x_i^k = \bar{x}_i$ and $\lim_{k \to \infty} \hat{\bs x}^k_{-i} = \bar{\bs x}_{-i}$ so that $(\bar{x}_i,\bar{\bs x}_{-i}) \in \Omega$ directly imply that $\lim_{k \to \infty} \ell_i(x_i^k, \hat f_i(\hat{\bs x}^k_{-i}, \tilde\theta_i)) = v > 0$, for any $\tilde\theta_i \notin \mc A_i$, while for all $\tilde\theta_i \in \mc A_i$ instead, $\lim_{k \to \infty} \ell_i(x_i^k, \hat f_i(\hat{\bs x}^k_{-i}, \tilde\theta_i)) = 0$. 
Since we are assuming that $\lim_{k \to \infty} \theta_i^k = \tilde{\theta}_i$, by considering a generic $\theta_i^k$ obtained from \eqref{eq:param_update} we know that for any $\tilde \varepsilon > 0$ there exists some $\tilde M \eqdef \tilde M(\tilde \varepsilon) \ge 0$ so that $|\ell_i(x_i^k, \hat f_i(\hat{\bs x}^k_{-i}, \theta^k_i))-v| \le \tilde \varepsilon$, i.e., $-\tilde \varepsilon + v \le \ell_i(x_i^k, \hat f_i(\hat{\bs x}^k_{-i}, \theta^k_i)) \le \tilde \varepsilon + v$, for all $k \ge \tilde M$.
On the other hand, we also know from $\lim_{k \to \infty} \ell_i(x_i^k, \hat f_i(\hat{\bs x}^k_{-i}, \tilde\theta_i)) = 0$ that for any $\bar \varepsilon > 0$ there exists some $\bar M \eqdef \bar M(\bar \varepsilon) \ge 0$ so that $\ell_i(x_i^k, \hat f_i(\hat{\bs x}^k_{-i}, \tilde\theta_i)) \le \bar \varepsilon$ for all $k \ge \bar M$. 

Without loss of generality, let us then pick $\tilde \varepsilon = v/2$ so that $-\ell_i(x_i^k, \hat f_i(\hat{\bs x}^k_{-i}, \theta^k_i)) \le -v/2$, and fix $\bar \varepsilon < \tilde \varepsilon = v/2$, which is always possible by continuity and $k$ sufficiently large. Under this latter condition, precisely $k > \textrm{max}\{\tilde M, \bar M\}$, evaluating the summation of the difference of the loss functions in $\bar \theta_i$ and $\theta^k_i$ at iteration $k$, with each term $\Delta \ell_{i,k}^t \eqdef \ell_i(x_i^t, \hat f_i(\hat{\bs x}^t_{-i}, \tilde\theta_i)) - \ell_i(x_i^t, \hat f_i(\hat{\bs x}^t_{-i}, \theta^k_i))$, leads to:
\ifTwoColumn
\else
$$
\frac{1}{k} \sum_{t = 1}^k \Delta \ell_{i,k}^t = \frac{1}{k} \left[\sum_{t = 1}^{\textrm{min}\{\tilde M, \bar M\} -1} \Delta \ell_{i,k}^t + \sum_{t = \textrm{min}\{\tilde M, \bar 	M\}}^{\textrm{max}\{\tilde M, \bar M\} -1} \Delta \ell_{i,k}^t + \sum_{t = \textrm{max}\{\tilde M, \bar M\}}^{k} \Delta \ell_{i,k}^t\right].
$$
\fi
While the first term in the summation is constant for fixed $\bar\varepsilon$ and $\tilde\varepsilon$ (and therefore for fixed $\bar M$ and $\tilde M$), namely $\sum_{t = 1}^{\textrm{min}\{\tilde M, \bar M\} -1} \Delta \ell_{i,k}^t \reqdef \alpha$, the second and third elements can always be upper bounded by $(k - \bar\rho)\bar \varepsilon - (k -\tilde\rho) v/2 + \beta$, where $\bar\rho$, $\tilde\rho \ge 0$ and $\beta \in \R$ are constants depending on the values $\tilde M$ and $\bar M$ take. Specifically, if $\bar M < \tilde M$, $\bar\rho = \bar M$, $\tilde\rho = \tilde M$ and $\beta < 0$, whereas if $\bar M \ge \tilde M$, $\bar\rho = \tilde M$, $\tilde\rho = \bar M$ and $\beta > 0$. Thus, taking the limit for $k \to \infty$ yields inequality:
$$
\lim\limits_{k \to \infty} \frac{1}{k} \sum_{t = 1}^k \Delta \ell_{i,k}^t \le \lim\limits_{k \to \infty} \frac{1}{k} [\alpha + \beta + (k - \bar\rho)\bar \varepsilon - (k -\tilde\rho) v/2] < 0
$$
in view of $\bar \varepsilon < \tilde \varepsilon = v/2$, meaning that we can always find some iteration index such that the cost obtained with some $\tilde\theta_i \in \mc A_i$ is strictly smaller than the one obtained with $\theta^k_i$ updated through the rule in \eqref{eq:param_update}, and assumed to be convergent to some $\tilde\theta_i \notin \mc A_i$. This clearly represents a contradiction and hence concludes the proof.
\hfill\qedsymbol

\smallskip

\emph{Proof of Proposition~\ref{prop:f}:}
Let us prove that $r(\bs x, \theta) = \sum_{i\in \mc N} \|
x_i-(\nu^{k+1}_i)^\top\bs x_{-i}-c^{k+1}_i\|^2_2$ satisfies Assumption~\ref{ass:convexity-1} and~\ref{ass:convexity-2}. 
Since $\Omega$ is a nonempty polytope, it is convex and bounded. Moreover, 
function $r$ is quadratic and positive semidefinite, and therefore continuous and convex \gls{wrt} $\bs x$, for all $\theta\in\R^p$. In addition, $r(\bs x, \theta)$ can be expressed also as:
\[
\begin{aligned}
	r(\bs x,\theta)&=\|A(\bs x)\theta-b(\bs x)\|_2^2\\ 
	&= \theta^\top A(\bs x)^\top A(\bs x)\theta+
	2b(\bs x)^\top A(\bs x)\theta+b(\bs x)^\top b(\bs x)
\end{aligned}
\]
where $A(\bs x)$, $b(\bs x)$ are suitably defined affine functions of $\bs x$, which
is a quadratic positive semidefinite function \gls{wrt} to $\theta$, and therefore convex
and differentiable for all $\bs x\in \Omega$.
Thus, 
\[
\begin{split}
	\underset{\bs x\in \Omega}{\textrm{sup}}~\left|\frac{\partial r(\bs x,\theta)}{\partial\theta}\right|&=
	\underset{\bs x\in \Omega}{\textrm{max}}~\left|2A(\bs x)^\top A(\bs x)\theta+A(\bs x)^\top b(\bs x)\right|\\
	&=\underset{i = 1,\ldots, q}{\textrm{max}}~\pm\{2A(v_i)^\top A(v_i)\theta+A(v_i)^\top b(v_i)\}
\end{split}
\]
where $v_1,\ldots,v_q$ are the vertices of $\Omega$ and the last inequality
follows since $2A(\bs x)^\top A(\bs x)\theta+A(\bs x)^\top b(\bs x)$ is convex and quadratic \gls{wrt} $\bs x$
and the maximum of a convex quadratic function over a polytope
is attained at one of its vertices. For all $\theta\in\R^p$, we thus obtain that:
\[  
\begin{split}
	&\underset{\bs x\in \Omega}{\textrm{max}}~\left|2A(\bs x)^\top A(\bs x)\theta+A(\bs x)^\top b(\bs x)\right|\\
	&=\max\{2A(v_{i(\theta)})^\top A(v_{i(\theta)})\theta+A(v_{i(\theta)})^\top b(v_{i(\theta)}),\\
	&~~~~~~~~-2A(v_{j(\theta)})^\top A(v_{j(\theta)})\theta+A(v_{j(\theta)})^\top b(v_{j(\theta)})\}\reqdef D(\theta)
\end{split}
\]
for some indices ${i(\theta)}$, ${j(\theta)}\in\{1,\ldots,q\}$.
By relying on Lemma~\ref{lemma:convergence} we hence obtain the desired result,
since all its assumptions are satisfied.
\hfill\qedsymbol

\smallskip

\emph{Proof of Theorem~\ref{th:convergence}:}
The convergence of each $\theta_i^k$ to some $\tilde{\theta}_i$ clearly implies that, in view of Standing Assumption~\ref{standing:learning_procedure}, also the associated composition mapping estimate converges, i.e., for all $i \in \mc N$, $\lim_{k \to \infty} \hat f_i(\bs x_{-i}, \theta^k_i) = \hat f_i(\bs x_{-i},\tilde{\theta}_i)$ with $\hat f_i(\bs x_{-i},\tilde{\theta}_i)$ possibly different from the true action-reaction mapping $f_i(\bs x_{-i})$. 
Then, in view of Proposition~\ref{prop:f}, we also have that $\lim_{k \to \infty} \hat{\bs x}^k = \tilde{\bs x}$. This latter relation in turn yields $\lim_{k \to \infty} \bs x^k = \bar{\bs x}$, as the action-reaction mappings $f_i(\cdot)$ are single-valued and continuous in view of Standing Assumption~\ref{standing:standard}. Note that both sequences $\{\hat{\bs x}^k\}_{k\in\N}$ and $\{\bs x^k\}_{k\in\N}$ are feasible in view of \eqref{eq:minimizers} and the fact that each $x_i^k$ is determined through $f_i(\cdot)$ that implicitly accounts for the common constraints $\Omega$.
Since $\hat{\bs x}^k$ is iteratively chosen following the tie-break rule induced by $\tfrac{1}{2}\|\cdot\|_2^2$, as the set of global minimizers in \eqref{eq:minimizers} may not be unique, it may happen that $\tilde{\bs x} \neq \bar{\bs x}$. We now make use of a local exact approximation argument to show that $\lim_{k \to \infty} \|\hat{\bs x}^k - \bs x^k\|_2 = 0$, and hence that the two limit points above coincide, and actually yield, by virtue of Lemma~\ref{lemma:local_exactness}, a stationary action profile $\bs x^\star$ in the sense of Definition~\ref{def:stat_action}.

In view of the definition of limit we can then always find some $\bar{k} \in \N$ and $\mu_i > 0$ such that a neighbourhood of $\bar{x}_i$, say $\mc B_{\mu_i}(\bar{x}_i)$, contains infinitely many points \cite{bachman2000functional}, meaning that $x_i^k \in \mc B_{\mu_i}(\bar{x}_i)$ for all $k \ge \bar{k}$, $i \in \mc N$. By virtue of the consistency property proved in Lemma~\ref{lemma:local_exactness} it thus follows that the pointwise approximation shall be exact, namely each $\tilde{\theta}_i$ is so that, for all $i\in\mc N$, $\| \hat f_i(\tilde{\bs x}_{-i},\tilde{\theta}_i) - f_i(\tilde{\bs x}_{-i})\|_2 = 0 = \| \hat f_i(\tilde{\bs x}_{-i},\tilde{\theta}_i) - \bar x_i\|_2$.

Since \eqref{eq:minimizers} is solved at the global optimum at every $k \in \N$, and the single-valuedness of each $f_i(\cdot)$, this latter relation readily yields $\sum_{i\in \mc N} \|\tilde x_i - \hat f_i(\tilde{ \bs x}_{-i}, \tilde \theta_i)\|^2_2 = \sum_{i\in \mc N} \|\tilde x_i - f_i(\tilde{ \bs x}_{-i})\|^2_2 = 0 = \sum_{i\in \mc N} \|\tilde x_i - \bar x_i\|^2_2$, and hence $\|\tilde x_i - \bar x_i\|_2 = 0$ for all $i \in \mc N$. This means that $\bar{ \bs x} = \tilde{ \bs x}$, and in view of Definition~\ref{def:stat_action}, they coincide with a fixed point of the action-reaction mappings $f_i(\cdot)$, as  $\sum_{i\in \mc N} \|f_i(\bar{\bs{x}}_{-i}) - \bar x_i\|^2_2=0$, i.e., $\|f_i(\bar{\bs{x}}_{-i}) - \bar x_i\|_2=0$ for all $i\in\mc N$, concluding the proof.
\hfill\qedsymbol

\subsection{Derivation of Kalman filter equations in \eqref{eq:KF}}\label{sec:kalman_eq}
The updates~\eqref{eq:KF} are obtained by applying linear Kalman
filtering to the linear time-varying system (we omit the index $i$
for simplicity of notation):
$\theta^{k+1}=\theta^k+\xi^k$, $x^k=(\phi^k)^\top\theta^k+\zeta^k$,
where $\xi^k$, $\zeta^k$  are zero-mean white noise terms with covariance $\beta I$ and 1, respectively.
The Kalman filter equations are:
\[
\begin{array}{rcl}
	M^k &=& P^k\phi^k/(1+(\phi^k)^\top P^k\phi^k)\\
	\theta^{k+\nicefrac{1}{2}} &=& \theta^{k} + M^k(x^k-(\phi^k)^\top\theta^k)\\
	P^{k+\nicefrac{1}{2}} &=& (I-M^k(\phi^k)^\top)P^k \\
	&=& P^k - P^k\phi^k(\phi^k)^\top P^k/(1+(\phi^k)^\top P^k\phi^k)\\
	\theta^{k+1} &=& \theta^{k+\nicefrac{1}{2}}\\
	P^{k+1}&=&P^{k+\nicefrac{1}{2}}+\beta I.
\end{array}
\]
Since $P^{k+\nicefrac{1}{2}}\phi^k = P^k\phi^k - P^k\phi^k(\phi^k)^\top P^k\phi^k/(1+(\phi^k)^\top P^k\phi^k)=P^k\phi^k/(1+(\phi^k)^\top P^k\phi^k) = M^k$,
we obtain $\theta^{k+1}=\theta^{k} + P^k\phi^k/(1+(\phi^k)^\top P^k\phi^k)(x^k-(\phi^k)^\top\theta^k)$.

\section*{Acknowledgment}
The authors thank Prof. Barbara Franci, Maastricht University, for the fruitful discussions on the technical proofs.


\bibliographystyle{IEEEtran}
\bibliography{learning_decision}



\end{document}